\documentclass[prb,twocolumn,superscriptaddress,showpacs,floatfix,amsmath,amssymb,preprintnumbers]{revtex4}
\usepackage{graphicx}

\newcommand{\beqn}{\begin{eqnarray}}
\newcommand{\eeqn}{\end{eqnarray}}
\newcommand{\eq}[1]{(\ref{#1})}
\newcommand{\dd}{{\mathrm{d}}}
\newcommand{\dual}{{{}^*}}
\newcommand{\Z}{{\mathbb Z}}
\newcommand{\dD}{{\mathrm D}}
\newcommand{\cZ}{{\mathcal Z}}

\begin{document}

\preprint{ITEP-LAT/2007-09}
\preprint{HU-EP-07/14}
\preprint{LU-ITP 2007/001}

\title{An Abelian two-Higgs model of strongly correlated electrons: \\
phase structure, strengthening of phase transition and QCD at finite density}

\author{M.~Bock}\affiliation{Institut f\"ur Theoretische Physik,
Universit\"at Leipzig, 04109 Leipzig, Germany}
\author{M.\,N.~Chernodub}
\affiliation{ITEP, B.Cheremushkinskaya 25, 117218 Moscow, Russia}
\author{E.-M.~Ilgenfritz}
\affiliation{Institut f\"ur Physik, Humboldt-Universit\"at zu Berlin,
Newtonstr. 15, 12489 Berlin, Germany}
\author{A.~Schiller}\affiliation{Institut f\"ur Theoretische Physik,
Universit\"at Leipzig, 04109 Leipzig, Germany}

\begin{abstract}
We investigate non-perturbative features of a three-dimensional Abelian Higgs model
with singly- and doubly-charged scalar fields coupled to a single compact Abelian
gauge field. The model is pretending to describe various planar systems of strongly
correlated electrons such as high-$T_c$ superconductivity in the overdoped regime and
exotic materials possessing excitations with fractionalized quantum numbers.
The complicated phase structure of the model is studied thoroughly using numerical
tools and analytical arguments. In the three-dimensional space of coupling
parameters we identify the Fermi liquid, the spin gap, the superconductor and the strange
metallic phases. The behavior of three kinds of topological defects -- holon and spinon
vortices and monopoles -- is explored in various phases. We also observe
a new effect, the strong enhancement of the phase transition strength reflected in
a lower order of the transition: at sufficiently strong gauge coupling the two second order phase
transitions -- corresponding to spinon-pair and holon condensation lines -- join partially
in the phase diagram and become a first order phase transition in that region.
The last observation may have an analogue in Quantum Chromodynamics at non-zero
temperature and finite baryon density. We argue that at sufficiently large baryon density
the finite-temperature transition between the (3-flavor paired) color superconducting phase
and the quark-gluon plasma phases should be much stronger compared with the transition between
2-flavor paired and 3-flavor paired superconducting phases.
\end{abstract}


\maketitle

\section{Introduction}

Gauge models involving multiple scalar fields coupled to an Abelian gauge field are applicable to a
large variety of systems such as multi-band superconductors~\cite{ref:two:component}, liquid metallic
hydrogen~\cite{ref:metallic}, easy-plane quantum antiferromagnets~\cite{ref:antiferromagnets}, {\it etc.}
These models have interesting phase structure and are distinguished by a copious zoo of topological defects.
Usually, all scalar (Higgs) fields in these models are considered to be alike so that they
are all equally charged and minimally coupled to the gauge field.
Contrary, in this paper we consider the two-Higgs model with a gauge field and two Higgs fields with
unequal charges. Our study is motivated by the fact that the charge-asymmetric two-Higgs gauge model can
emerge as an effective description of unconventional superconductivity~\cite{ref:review}.

Despite the two-Higgs model is formulated in a very simple way, it can actually capture basic properties
of various systems: instanton (monopole) plasma described by the compact Abelian gauge model,
superfluidity (the $XY$ spin model), the $\Z_2$-gauge model which has analogues in
particle physics and the coupled $\Z_2$-gauge-$XY$ model describing fractionally charged
excitations in strongly correlated electron systems. It shares also a similarity with
a Ginzburg-Landau model with two vector fields \cite{ref:UV:model} which was suggested to
describe extended $s$- and $d$-wave superconducting granular systems.

The common feature of all unconventional superconductors~\cite{ref:Anderson:introduction}
with high critical temperatures is the presence of copper oxide layers.
The layered structure is seen both in electronic~\cite{ref:review:Dagotto}
and optical~\cite{ref:Basov} anisotropic structure of the cuprates. The anisotropy is a key
ingredient of various approaches to this phenomenon~\cite{ref:review}. The CuO${}_2$ layers
are associated with conductance plates while the atoms in the space between the layers form
a so-called charge reservoir which supplies the charge carriers to the planes.
The charge reservoirs themselves are almost insulating in the superconducting phase~\cite{ref:Basov}.
The charge carriers can be either electrons or holes depending on the nature of
the dopant. The fraction of the carriers $p$ in the planes is controlled by the doping level $x$
which is usually encoded in the chemical formula of the cuprate oxides ({\it i.e.}, $p = x$
in the structurally simple La${}_{2-x}$Sr${}_x$ CuO${}_4$ as a prototype of many of
the cuprate materials). In the clean limit, $x=0$, the cuprates are Mott insulators,
while at certain $x$ the cuprate becomes a poor conductor which -- at low enough temperature --
turns into a superconductor. Nowadays, the critical temperatures have climbed the level of 140 K
in Hg-based cuprates.

We concentrate on the in-plane mechanism of the high-$T_c$ superconductivity restricting ourselves
to the slave-boson approach in the $t$-$J$ model. That model is used to describe the ground state of the high-$T_c$ superconductor~\cite{ref:Anderson,ref:slave-boson,ref:Baskaran:Solid:State}
as charge carriers (electrons or holes) in the two-dimensional copper-oxide plane.
The $t$-$J$ Hamiltonian~\cite{ref:Anderson} describes hopping holes (or electrons) and localized spins
in a plane:
\beqn
  H_{tJ} = - t \!\! \sum\limits_{<ij>,\sigma} \!\!
  c_{i\sigma}^\dagger P_{ij,-\sigma} c_{j\sigma}
  + J \!\sum\limits_{<ij>} ({\vec S}_i {\vec S}_j - \frac{1}{4} n_i n_j)
  \,.
  \label{eq:tJ}
\eeqn
The first term specifies holes (electrons) moving without flipping the spin $\sigma$.
Double occupancy is explicitly forbidden by the presence of the projectors
\[
  P_{ij,\sigma} = (1 - n_{i,\sigma})(1 - n_{j,\sigma})\,.
\]
The second term describes the anti-ferromagnetic Heisenberg coupling between spins located at the copper sites. Here
\[
  \vec S_i= \frac{1}{2} \sum_{\sigma\sigma'}c_{i\sigma}^\dagger {\vec \tau}_{\sigma\sigma'}  c_{i\sigma'}
\]
is the spin operator, $c^{\dagger}_{i\sigma}$, $c_{i\sigma}$ are the hole (or electron) creation and annihilation operators, and
\[
  n_{i,\sigma} = c_{i\sigma}^\dagger c_{i\sigma}\,,
  \qquad
  n_i= n_{i,\uparrow} + n_{i,\downarrow}
\]
denotes the occupation numbers.

According to a popular scenario~\cite{ref:slave-boson}, the electron degree of freedom can be split
into the spin and charge constituents (spinon and holon, respectively). The splitting gives rise to an internal gauge degree of freedom, with respect to which the spinons and holons have positive and negative charges (say, $+1$ and $-1$, respectively). The internal group is necessarily compact and this leads to a specific interaction between the spin-charge separated constituents to be discussed later.

Under certain conditions the spinon particles get paired and form pairs similarly to Cooper pairs
in ordinary superconductivity. Then the pairs of spinons are presented by a spinon-pair field which has
charge $+2$ with respect to the internal group. Therefore, in a mean field approach, the system is
described by two scalar (Higgs) fields: the holon and the spinon-pair field with internal charges $-1$ and $+2$, respectively. Both kind of fields interact via the exchange of an internal gauge field which is compact by construction.

It is important to note that the group for the internal gauge degree of freedom has nothing to do with
the usual Maxwell electromagnetic group. For example, the internal degree is compact while the electromagnetism is described by a non-compact Abelian group. The original spinon is an electromagnetically neutral excitation while the holon is the only constituent which carries the electric charge.
Thus the charge carrier is the holon while the spinon may affect the properties of the strongly correlated
material only indirectly: the formed spinon-pairs interact via the compact gauge field with the holons.

Entering a more technical description, the creation operators are decomposed
as~\cite{ref:slave-boson,ref:Baskaran:Solid:State}
\beqn
  c^\dagger_{i\sigma} = f^\dagger_{i\sigma} b_i \, ,
\eeqn
with the constraint
\beqn
  f^\dagger_{i\uparrow} f_{i\uparrow} + f^\dagger_{i\downarrow} f_{i\downarrow} + b^\dagger_{i} b_{i}
  = 1 \, .
\eeqn
Here $f_{i\sigma}$ is a spin-particle (``spinon'') operator and $b_{i}$ a charge-particle (``holon'') operator.

In addition to the ordinary electromagnetic (external) gauge symmetry,
\beqn
  U(1)_{\mathrm{ext}}:\ \, \
  c_{i\sigma} \to e^{i \omega} c_{i\sigma},\ \,
  f_{i\sigma} \to f_{i\sigma}, \ \,
  b_{i} \to e^{i \omega_i} b_{i\sigma}
  \label{eq:gauge:exernal}
\eeqn
the spin-charge separation naturally introduces an (internal) compact $U(1)$ gauge freedom,
\beqn
  U(1)_{\mathrm{int}}:\ \, \
  c_{i\sigma} \to c_{i\sigma},\ \,
  f_{i\sigma} \to e^{i \alpha_i} f_{i\sigma}, \ \,
  b_{i} \to e^{i \alpha_i} b_{i\sigma}
  \label{eq:gauge:internal}
\eeqn
which plays an essential role~\cite{ref:Baskaran,ref:Larkin} in understanding the physics
of strongly correlated electrons. The spin-charge separation idea may also be applied to various systems
including the general case of non-relativistic electrons~\cite{ref:Pauli} as well as the case of strongly interacting gluons in Quantum Chromodynamics~\cite{ref:slave-boson-further}.

The effective theory of superconductivity can further be simplified and reformulated in terms of lattice gauge models~\cite{ref:Baskaran,ref:Larkin,ref:LeeNagaosa:characterization,NagaosaLee,ref:Ichinose:Matsui},
see recent reviews~\cite{ref:Baskaran:review}.
Thus, the $t$-$J$ model~\eq{eq:tJ} is related to a compact Abelian gauge model with the internal symmetry~\eq{eq:gauge:internal}, which couples holons and spinons.
As in usual BCS superconductivity, under appropriate conditions the spinons couple and form bosonic quasiparticles. In a mean field theory one can define fields which behave under the gauge transformations~\eq{eq:gauge:internal} as:
\beqn
  \chi_{ij} & = & \sum\nolimits_\sigma \langle f^\dagger_{i\sigma} f_{j\sigma} \rangle
  \to \chi_{ij} \cdot
  {\mathrm e}^{- i (\alpha_i - \alpha_j)}\,, \\
  \Delta_{ij} & = & \langle  f_{i\uparrow}f_{j\downarrow}
                       - f_{i\downarrow}f_{j\uparrow}  \rangle
  \to \Delta_{ij} \cdot {\mathrm e}^{i (\alpha_i + \alpha_j)}\,.
\eeqn
The phase of the field $\chi$ represents nothing but the compact $U(1)$ gauge field,
\[
  \theta_{ij} \equiv \arg \chi_{ij} \to \theta_{ij} + {(\dd \alpha)}_{ij}
\]
with ${(\dd \alpha)}_{ij} = \alpha_j - \alpha_i$, and the radial part,
$\chi = | \langle \chi_{ij} \rangle |$, is the so-called ``resonating valence bond'' (RVB) coupling.
The doubly-charged spinon-pair field $\Delta$ is analogous to the Cooper pair.

At high temperature the RVB coupling vanishes, $\chi=0$, and the system is in the Mott insulator
(or ``poor metallic'') phase. With decreasing temperature $\chi$ acquires a non-zero value, eventually enabling the formation of a spinon-pair condensate $\Delta = |\langle\Delta_{ij}\rangle|$
and/or of a  holon condensate $b = \langle b_i\rangle$~\cite{ref:Baskaran:Solid:State}.
Therefore, four phases~\cite{ref:LeeNagaosa:characterization,NagaosaLee,ref:Ichinose:Matsui} may emerge:
\begin{itemize}
  \item[(i)]   the Fermi liquid (FL) phase with $b \ne 0$, $\Delta=0$;
  \item[(ii)]  the spin gap (SG) phase with $b =0$, $\Delta\ne 0$;
  \item[(iii)] the superconductor (SC) phase with $b \ne 0$, $\Delta\ne 0$;
  \item[(iv)]  the strange metallic (SM) phase with $b =0$, $\Delta= 0$.
\end{itemize}

\begin{figure}[!htb]
  \begin{center}
  \includegraphics[width=7.5cm]{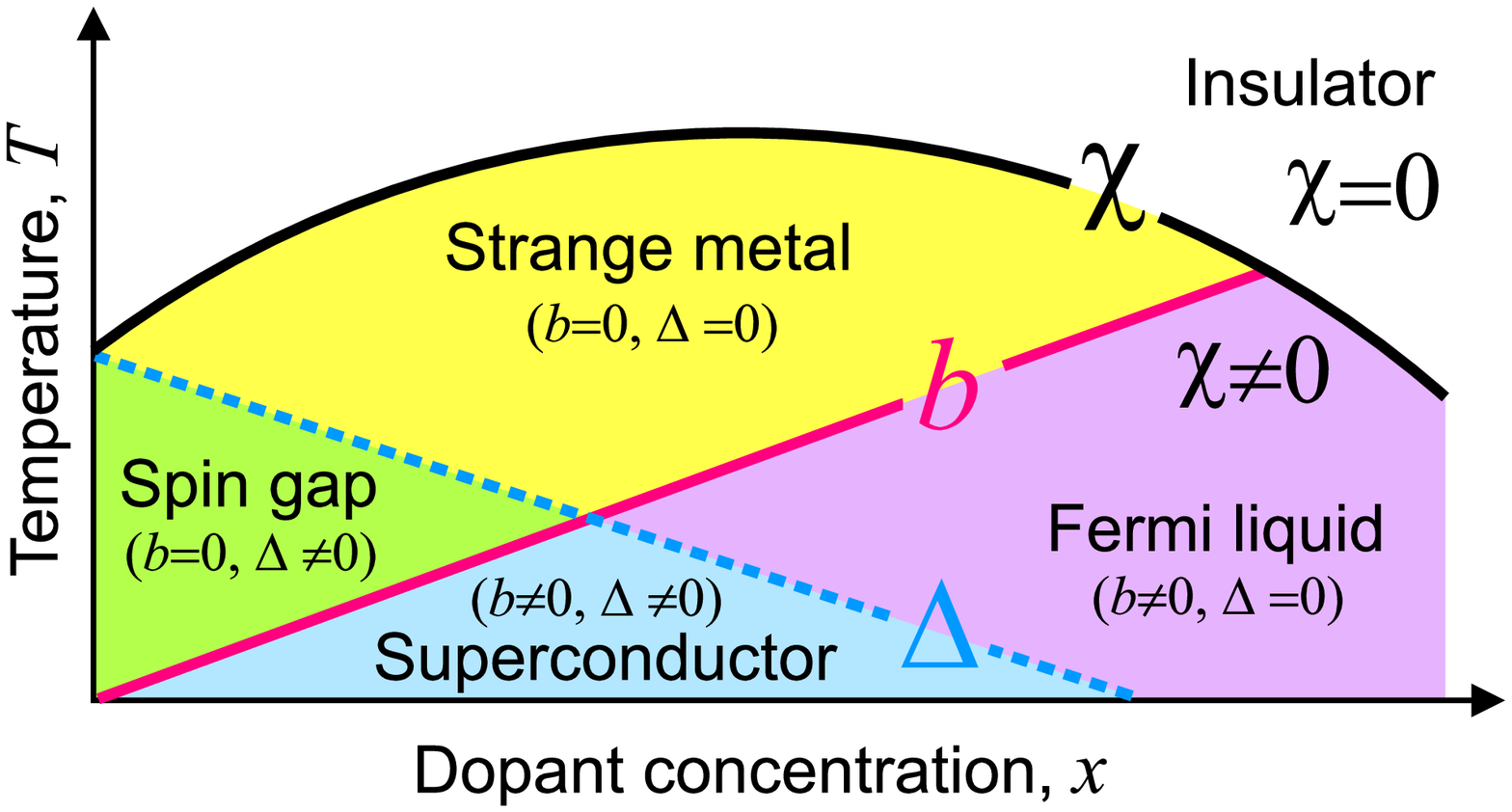}
  \end{center}
  \caption{The phase diagram in the plane ``temperature-dopant concentration'' (suggested
           in Ref.~\onlinecite{ref:LeeNagaosa:characterization}).}
  \label{fig:Lee:Nagaosa}
\end{figure}

The ground state of the superconducting layer is proposed~\cite{ref:LeeNagaosa:characterization,NagaosaLee} to be described by a compact Abelian two-Higgs model (cA2HM) in {\it three} dimensions with a $U(1)$ gauge link field $\theta$, a singly-charged ($q_h=1$) holon field $\Phi_h$, and a doubly-charged ($q_s=2$)
spinon-pair field $\Phi_s$. More precisely, the classical three-dimensional statistical model describes
the ground state of the zero temperature {\it two}-dimensional spin-charge separated quantum system
of holons and spinons coupled together by the internal compact gauge field. The model can be considered as a phenomenological extension of the Ginzburg-Landau model for ordinary low-$T_c$ superconductors.

The model is applicable to the overdoped regime of high-$T_c$ materials (called sometimes the SM
region~\cite{ref:review,ref:strange:metal}) where the $SU(2)$ particle-hole symmetry is explicitly broken. Further arguments for justifying this approach and a discussion of its limitations can be found in Ref.~\onlinecite{ref:review}.

Since high-$T_c$ materials are type-2 superconductors, we restrict ourselves to the London limit in which the radial parts of both Higgs fields $\Phi_k = |\Phi_k| \, e^{i \varphi_k}$, $k=h,s$ are frozen, $|\Phi_{h,s}| = {\mathrm{const}}$. The action of this compact Abelian two-Higgs model is
\beqn
  & & S_{\mathrm{cA2HM}}
  = -\beta \sum\limits_P \cos\theta_P \nonumber\\
  & & - \kappa_h \sum\limits_l \cos({\mathrm{d}} \varphi_h + \theta)_l
  - \kappa_s \sum\limits_l \cos({\mathrm{d}} \varphi_s + 2 \theta)_l\,,
  \label{eq:cA2HM:action}
\eeqn
where
$\theta_P \equiv (\dd \theta)_P$ is the standard lattice plaquette.

The model~\eq{eq:cA2HM:action} obeys a lattice version of the $U(1)$ internal gauge symmetry~\eq{eq:gauge:internal}:
\beqn
  \theta \to \theta + {\mathrm{d}} \alpha\,, \quad \varphi_k \to \varphi_k - q_k \alpha\,,
\eeqn
with $k=h,s$. It describes the hole (electron) ``constituents'' by the dynamical holon $\varphi_h$ and spinon $\varphi_s$ phases, which strongly interact via the dynamical gauge field $\theta$. The inverse gauge coupling $\beta$ in Eq.~\eq{eq:cA2HM:action}, $\beta = \chi_f + \chi_b$, is given by the diamagnetic susceptibilities $\chi_f$ of the spinon and $\chi_b$ of the holon fields.
The hopping parameters $\kappa_k$ are connected to the doping concentration $x$ and the couplings $t$ and $J$ of \eq{eq:tJ} as follows: $\kappa_h \propto t \cdot x$ and $\kappa_s \propto J$
(see~\onlinecite{ref:review}).

The phase diagram of the model~\eq{eq:cA2HM:action} and the basic properties of the topological defects
were studied for a limited set of coupling parameters in our preliminary investigation~\cite{ref:PRB2006}.
The aim of the present paper is to extend the study to a much larger coupling range using both numerical
and analytical tools. We also identify the order of various phase transitions and confirm the existence of a phase transition strengthening effect which was first suggested in Ref.~\onlinecite{ref:PRB2006}.
The enhancement is reflected in decreasing the order of  the phase transition at a common joint segment of two second order transitions. The signatures of unexpected strengthening of the phase transition were later observed in a different model describing a gauge field coupled to two Higgs fields of equal charges~\cite{ref:Kragset}.

The structure of the paper is as follows. In Section~\ref{sec:topdefects} we show that the model contains three types of topological defects: two types of vortices and one type of monopoles. We discuss the simplest features of the topological defects and derive their effective action using analytical tools only. In Section~\ref{sec:phasediagram} we describe the phase diagram of the model in the three-dimensional space of gauge, holon and spinon couplings using known results available for less complicated systems. The limiting cases of the three-dimensional phase-cube are analyzed in detail and the possible structure of the phase transition surfaces in the 3D-coupling space is pointed out. Section~\ref{sec:numerical} is devoted to a numerical investigation of the phase diagram by Monte-Carlo simulations. We analyze various two-dimensional cross-sections of the 3D-phase diagram identifying numerically the order of the phase transitions. The behavior of thermodynamical quantities in combination with that of the densities of the topological defects allows us to identify the nature of the phases in different regions of the 3D-coupling space. In the same Section we confirm the effect of the phase transition strengthening due to merging transition lines. The phase transition enhancement may also be relevant for Quantum Chromodynamics which describes the theory of strong interactions. We point out in Section~\ref{sec:QCD} that the phase diagrams of the cA2HM and QCD contain common features including the joining of transition lines. We suggest that QCD at high temperature and high baryon density may experience the same strengthening effect. The last Section is devoted to our conclusions.

\section{Properties of topological defects}
\label{sec:topdefects}

The model~\eq{eq:cA2HM:action} contains three kinds of topological defects: a monopole and two types of vortices, referred to as the holon and the spinon vortex~\cite{ref:review,ref:LeeNagaosa:characterization}.
The monopole has magnetic charge $2 \pi \hbar$ while the holon (spinon) vortex carries magnetic flux quanta $2 \pi \hbar$ ($\pi \hbar$) of the gauge field $\theta$. One monopole is simultaneously a source of one holon vortex and two spinon vortices. In this Section we discuss some basic properties of these defects.

Despite the model~\eq{eq:cA2HM:action} is formulated in the Wilson representation (with an action of cosine-type), the basic properties of the topological defects can be guessed from the so-called Villain representation~\cite{ref:Villain} which is more suitable for analytical considerations. A similar set of transformations was performed in a non-compact model in Ref.~\onlinecite{ref:two:component}.
The principal difference between the compact (considered here) and non-compact models~\cite{ref:two:component} is the presence of monopoles, and, as a consequence, a richer phase diagram due to the existence of confining phases.

The Villain representation of the partition function of the model~\eq{eq:cA2HM:action} is
\beqn
  & & \cZ = \int\limits^\pi_{-\pi} \dD \theta \sum_{n \in \Z(c_2)} \,
  \int\limits^\pi_{-\pi} \dD \varphi_h \int\limits^\pi_{-\pi} \dD \varphi_s
  \sum_{l_h \in \Z(c_1)} \sum_{l_s \in \Z(c_1)} \, \nonumber\\
  & & \exp\Bigl\{ - \beta {||\dd \theta + 2 \pi n||}^2
  - \tilde\kappa_h {||\dd \varphi_h + \theta + 2 \pi l_h||}^2 \nonumber\\
  & & \hspace{32mm} - \tilde\kappa_s {||\dd \varphi_s + 2 \theta + 2 \pi l_s||}^2\Bigr\},
  \label{eq:Z:initial}
\eeqn
where the Villain couplings $\tilde\kappa_h$ and $\tilde\kappa_s$ correspond to the Wilson couplings
$\kappa_h$ and $\kappa_s$. There are three integer-valued forms: the plaquette form $n$ and two link forms $l_h$ and $l_s$.

The definition~\eq{eq:Z:initial} is written in a convenient condensed form by means of differential forms on the lattice~\cite{ref:differential}. In brief, the notations are as follows. Let $a$ and $b$ be two $r$-forms on the lattice. Here $r=0$ corresponds to scalars (with the support on sites, $c_0$), $r=1$ corresponds to vectors (with the support on links, $c_1$), {\it etc}. Then the scalar product $(a,b)$ is defined as a scalar product over the whole support of the $r$-form (sites, links etc.) over the lattice.
Thus, for two vector forms (``one-forms''), we have $(a,b) = \sum_l a_l b_l$. The modulus squared is then defined as $||a||^2 \equiv (a,a)$. The finite-difference operator ``d'' increases the rank of the form by one, $r \to r + 1$ (thus having the meaning of a gradient), while the operator $\delta \equiv \dual \dd \dual$ (to be used below) decreases the rank of the form, $r \to r - 1$ (having the meaning of a divergence). The duality operator -- which switches forms between the original and the dual lattices (by element-wise equating the values assigned to the dual to each other supports) -- is denoted as ``$*$''.

The Berezinskii-Kosterlitz-Thouless (BKT) transformation~\cite{ref:BKT} allows us to rewrite the partition function~\eq{eq:Z:initial} in terms of monopoles and vortices. The monopoles appear due to the compactness of the gauge fields $\theta$, and two types of vortices arise from the presence of two independent species of Higgs fields. The monopole ``trajectory'' (in $3D$ actually located on cubes) is denoted as $j$,
the vortex ``trajectory'' (in $3D$ located on plaquettes) are $\sigma_h$ (the holon vortex) and $\sigma_s$ (the spinon vortex). Using the standard approach we represent the integer-valued  form $n$ as a sum
$n = \dd q + m[j]$ of the co-closed surface $\dd q$ and the surface $m[j]$ spanned on the monopole trajectory $j$: $\dual j = \delta \dual m[j]$, or, equivalently, $j = \dd m[j]$. The two-form $n$ can also be represented in Hodge-de-Rham form as a sum of a closed and a co-closed part
\beqn
  n = \delta \Delta^{-1} j + \dd (\Delta^{-1} \delta m[j] + s)\,,
  \label{eq:n:expansion}
\eeqn
where $\Delta = \dd \delta + \delta \dd$ is the lattice Laplace operator, $\Delta^{-1}$ is its inverse,
and $s \in \Z(c_1)$, $m \in \Z(c_2)$, and $j \in \Z(c_3)$ are integer-valued forms. The monopole current
is closed
\beqn
  \delta \dual j = 0\,.
  \label{eq:j:closed}
\eeqn

Substituting Eq.~\eq{eq:n:expansion} into the first term under the exponent in Eq.~\eq{eq:Z:initial} we get
\beqn
  \dd \theta + 2 \pi n = \dd A + 2 \pi \delta \Delta^{-1} j\,,
  \label{eq:d:theta}
\eeqn
where
\beqn
  A = \theta + 2 \pi (\Delta^{-1} \delta m[j] + s)
\eeqn
is a non-compact gauge field.

The same trick can be performed with the phases $\varphi_h$ and $\varphi_s$ (in the second and the third term in the exponent in Eq.~\eq{eq:Z:initial})
\beqn
  l_k = \delta \Delta^{-1} \sigma^{(0)}_k + \dd (\Delta^{-1} \delta p_k[\sigma^{(0)}_k] + r_k)\,,
  \label{eq:lq:expansion}
\eeqn
where $\sigma^{(0)}_k \in \Z(c_2)$ is the {\it closed} part of the holon (if $k=h$) or spinon (if $k=s$)
vortex trajectory,
\beqn
  \delta \dual \sigma^{(0)}_k \in \Z(c_2) = 0\,, \quad k=h,s\,,
\eeqn
$p_k \in \Z(c_2)$, and $r_k \in \Z(c_0)$. The analogue of Eq.~\eq{eq:d:theta} can be written in the following form
\beqn
  \dd \varphi_k + q_k\, \theta + 2 \pi l_k = \dd \varphi^{\mathrm{n.c.}}_k + q_k A +
   2 \pi \delta \Delta^{-1} \sigma_k\,.
  \label{eq:d:varphi}
\eeqn
Here the vortex trajectories are given by the combinations
\beqn
  \sigma_k = \sigma^{(0)}_k - q_k \, m[j]\,,
\eeqn
and the non-compact Higgs phases are
\beqn
  \varphi^{\mathrm{n.c.}}_k = \varphi_k + 2 \pi (\Delta^{-1} \delta p_k[\sigma^{(0)}_k] + r_k)\,.
\eeqn

\begin{figure}[!htb]
  \begin{center}
  \includegraphics[width=5.0cm]{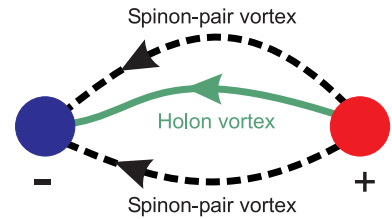}
  \end{center}
  \caption{An example of an monopole-vortex configuration.}
  \label{fig:mon:conf}
\end{figure}

The vortex trajectories end on the monopole ``trajectories'', which are points (instantons) in three-dimensional space-time,
\beqn
  \delta \dual \sigma_k + q_k \dual j = 0\,, \qquad k=h,s\,.
  \label{eq:conservation}
\eeqn
The above equation means that there are one holon vortex and two spinon vortices -- each carrying a corresponding elementary flux -- attached simultaneously to each monopole. A typical configuration monopole-vortex configuration is shown in Figure~\ref{fig:mon:conf}.

Substituting Eqs.~\eq{eq:d:theta} and \eq{eq:d:varphi} into \eq{eq:Z:initial} and performing the Gaussian
integration over the non-compact fields $A$, $\varphi^{\mathrm{n.c.}}_h$, and $\varphi^{\mathrm{n.c.}}_s$,
one gets the BKT representation of the two-Higgs model:
\beqn
  \cZ = \sum_{\dual j \in \Z(\dual c_3)} \,
  \sum_{\stackrel{\dual \sigma_{h,s} \in \Z(\dual c_2)}{\delta \dual \sigma_k + q_k\, \dual j = 0}} \!\!\!
  e^{- S^{\mathrm{mon}}[j] - S^{\mathrm{vort}}[\sigma_h,\sigma_s]}\,,
  \label{eq:Z:BKT}
\eeqn
where an irrelevant constant factor is omitted. The monopole and the vortex actions are, respectively,
\beqn
  S^{\mathrm{mon}}[j]  & = &  4 \pi^2 \beta \Bigl(j, \frac{1}{\Delta + m_\gamma^2} j \Bigr)\,,
  \label{eq:S:mon}\\
  S^{\mathrm{vort}}[\sigma_h,\sigma_s] & = &  4 \pi^2 \sum_{k,k' = h,s}
  \Bigl(\sigma_k, {\mathcal K}_{kk'} \, \sigma_{k'} \Bigr)
  \label{eq:S:vort}
\eeqn
with the differential operator ${\mathcal K}$
\beqn
  {\mathcal K}_{kk'}  = \frac{\tilde\kappa_{k}}{\Delta}
  \Bigl(\delta_{kk'} - \frac{q_k}{q_{k'}} \frac{\omega_{k'}\,
   m_\gamma^2}{\Delta + m_\gamma^2}\Bigr)\,.
  \label{eq:K}
\eeqn
In Eqs.~\eq{eq:S:mon}, \eq{eq:K} $m_\gamma$ denotes the mass of the photon,
\beqn
  m_\gamma^2 = \frac{1}{\beta} (\tilde\kappa_h + 4 \tilde\kappa_s)\,,
  \label{eq:photon:mass}
\eeqn
and $\omega_k$ are the normalized weights
\beqn
  \omega_k = \frac{q^2_k \, \tilde\kappa_k}{\tilde\kappa_h + 4 \tilde\kappa_s}\,, \qquad
  \omega_h + \omega_s = 1\,.
  \label{eq:weight}
\eeqn

The (self-)interaction of the monopoles and vortices can readily be read off from Eqs.~\eq{eq:Z:BKT},
\eq{eq:S:mon} and \eq{eq:S:vort}. The monopoles interact via a massive photon exchange, and
thus the interactions of the monopole ``trajectories'' are suppressed at large distances.

The vortex interactions contain a long-range term. The origin of the long-range forces is simple:
in the Abelian model with one Higgs field the massless Goldstone mode is eaten up by the longitudinal
component of the gauge field. Thus, the gauge field becomes massive, whereas the massless Goldstone boson disappears. This is not the case in the present model with two Higgs fields: one of the Goldstone bosons can be absorbed into the longitudinal component of the gauge field, while the other remains alive.
Thus, the two-Higgs system with one gauge field will always have one massless excitation which, in particular, leads to a long-range interaction between the vortex ``surfaces''.

The vortex-vortex interaction term~\eq{eq:S:vort}, \eq{eq:K} can also be rewritten in the form
\beqn
  S^{\mathrm{vort}}[\sigma_h,\sigma_s] = S^{\mathrm{vort}}_{hh}[\sigma_h]
  + S^{\mathrm{vort}}_{ss}[\sigma_s] + S^{\mathrm{vort}}_{sh}[\sigma_h,\sigma_s]
  \label{eq:vortex:int}
\eeqn
where the holon-holon, spinon-spinon and holon-spinon vortex interactions are, respectively,
\beqn
  && \hskip -10mm
  S^{\mathrm{vort}}_{hh} = 4 \pi^2 \tilde\kappa_h \biggl(\sigma_h,\Bigl[\omega_h \cdot
  \frac{1}{\Delta + m_\gamma^2}
  + \omega_s \cdot \frac{1}{\Delta}\Bigr] \sigma_h \biggr)\,,
  \label{eq:vort:hh}\\
  && \hskip -10mm
  S^{\mathrm{vort}}_{ss} = 4 \pi^2 \tilde\kappa_s \biggl(\sigma_s, \Bigl[\omega_s \cdot
  \frac{1}{\Delta + m_\gamma^2}
  + \omega_h \cdot \frac{1}{\Delta}\Bigr] \sigma_s \biggr)\,,
  \label{eq:vort:ss}\\
  && \hskip -10mm
  S^{\mathrm{vort}}_{hs} = 8 \pi^2 {(\tilde\kappa_h \tilde\kappa_s\,\omega_h \omega_s)}^{1/2}
  \biggl(\sigma_h, \Bigl[\frac{1}{\Delta + m_\gamma^2}- \frac{1}{\Delta}\Bigr]\sigma_s\biggr).
  \label{eq:vort:sh}
\eeqn
{}From these expressions it can be seen that the short-range interaction between the parallel segments
of all vortices is always repulsive. However, the presence of the long-range component acts in a
different way: segments of equal type vortices are repulsive (holon-holon and spinon-spinon),
while the parallel vortex segments of different types (holon-spinon) are always attractive.
The respective strengths of the repulsion and attraction depend on the weights of the holon and spinon vortices, Eq.~\eq{eq:weight}, which, in turn, depend on the strengths of the holon and spinon-pair condensates, $\tilde\kappa_k \propto {|\Phi_k|}^2$.

The presence of the massless mode in the interaction between vortices \eq{eq:vort:hh}, \eq{eq:vort:ss},
\eq{eq:vort:sh} on the Lagrangian level does not mean that on the quantum level the interactions remain
unscreened. In contrary, we expect that the massless mode should disappear due to non-perturbative effects.
This happens, for example, in the monopole gas of the three-dimensional compact Abelian model~\cite{ref:Polyakov}. Indeed, in this example the bare interaction between the monopoles is of the Coulomb type, $(j, \Delta^{-1} j)$, while all correlations in the statistical ensembles of the monopoles are exponentially suppressed at large distances by a Debye mass which appears due to monopole interactions in the plasma regime. A similar effect is expected in the considered system of Coulomb-interacting vortices. In the dilute ensemble of vortices the screening mass is expected to be small compared to the mass of the photon $m$, Eq.~\eq{eq:photon:mass}, but still non-zero. Therefore one may expect that on the quantum level the Goldstone mode may disappear.

\section{The phase diagram}
\label{sec:phasediagram}

The internal structure of the three-dimensional phase diagram of the (London limit version of the) compact
Abelian two-Higgs model~\eq{eq:cA2HM:action} is rather complicated. However, the faces and the edges of the cube representing the ``compactified'' phase diagrams can be drawn relatively easily because they are related to various well-known condensed-matter systems. These limiting cases of the phase cube correspond to appropriate combinations of vanishing and/or infinitely large couplings $\beta$, $\kappa_h$ and $\kappa_s$. Below we discuss these limits in detail.

\subsection{The $(\beta,\kappa_h)$-faces}

The $\kappa_s$ parameter defines the coupling of the spinon-pair field to the compact gauge field.
Below we consider two limiting cases corresponding to vanishing and infinitely strong coupling $\kappa_s$.

\subsubsection{The $\kappa_s=0$ face: $Q=1$ compact Abelian Higgs model}

The $\kappa_s=0$ face of the cA2HM corresponds to the $Q=1$ Higgs model with a compact Abelian field (cAHM${}_{Q=1}$) in three dimensions:
\beqn
  S_{\mathrm{cAHM}_{Q=1}} = -\beta \sum\limits_P \cos\theta_P -
  \kappa_h \sum_l \cos({\mathrm{d}} \varphi_h + \theta)_l \, .
  \label{eq:cAHM1}
\eeqn
On that face the holon condensate is coupled to the compact $U(1)$ gauge field while the spinon-pair field is decoupled from all other fields. The phase of the spinon-pair field is disordered which implies condensation of the spinon vortices and, therefore, vanishing of the spinon-pair condensate $\Delta$.

\begin{figure}[!htb]
  \begin{center}
  \includegraphics[width=6.0cm]{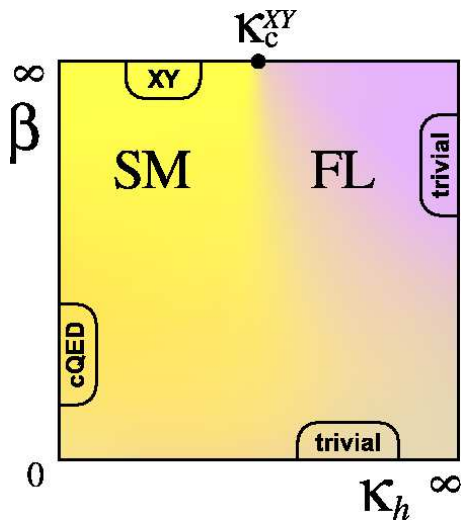}
  \end{center}
  \caption{The phase diagram on the $\kappa_s=0$ face corresponds to the compact Abelian $Q=1$ Higgs model
           with the action~\eq{eq:cAHM1}. The SM and FL phases in the interior of the diagram, as well as
	   the limiting models ($XY$, cQED and the trivial cases) are explicitly indicated. The
	   $XY$-critical point is given in Eq.~\eq{eq:XY:critical}.}
  \label{fig:phase:2D:I}
\end{figure}

The cAHM${}_{Q=1}$ has extensively been studied in the literature and it seems that a consensus on the phase structure is reached~\cite{ref:fradkin,Osterwalder:1977pc,ref:einhorn,ref:AHM1}.
The phase diagram in the $(\beta$,$\kappa_h)$-plane is plotted in Figure~\ref{fig:phase:2D:I}.

It contains two phases:
\begin{itemize}
  \item[(i)]   the SM phase at small $\beta$/small $\kappa_h$,
  \item[(ii)]  the FL phase at large $\beta$/large $\kappa_h$.
\end{itemize}
The FL phase is the broken/Higgs phase with nonzero holon condensate $b \neq 0$, and the SM phase is the confining/symmetric phase with $b\approx 0$. The distinction between these phases is blurry since the broken phase is always partially confining while (at the opposite end) even in the deeply confining phase
traces of the Higgs condensate can be found. The usual order parameter -- the Higgs (holon) condensate -- is, strictly speaking, non-zero in the whole phase diagram, and thus it is usually said that these phases are ``analytically connected''. As a consequence there is no {\it local} order parameter in terms of the primary fields of the model, which could in principle discriminate between the phases.

Despite the boundary between broken and confining ``phases'' has definitely not the characteristics of a phase transition in the thermodynamic sense, these phases may be discriminated by the condensation properties of topological defects. A non-thermodynamic boundary of this type is known in the literature as a Kert\'esz line~\cite{ref:Kertesz} which is defined as a line where the vortices start/cease to condense.
This line -- which is not plotted in Figure~\ref{fig:phase:2D:I} -- has been thoroughly studied in the three-dimensional Abelian Higgs model in Ref.~\onlinecite{Wenzel:2005nd} and was suggested to appear in
models of particle physics as well\cite{ref:particle:physics}. Note that in the present context the word ``line'' refers to a two-dimensional coupling parameter space, while in, say, a three-dimensional parameter space the corresponding manifold is a surface.

The edges of the cube in the $\kappa_s=0$-plane can be analyzed following Refs.~\onlinecite{ref:fradkin,Osterwalder:1977pc,ref:einhorn,ref:AHM1}. At the $\kappa_h =0$ edge the model is basically a plasma model. In fact, along the $\kappa_h = \kappa_s=0$ edge the cA2HM reduces to the pure compact $U(1)$ gauge theory,
\beqn
  S_{\mathrm{cQED}} = -\beta \sum\limits_P \cos\theta_P\,,
  \label{eq:cQED}
\eeqn
which is known to be confining at any value of $\beta$ due to the presence of monopoles~\cite{ref:Polyakov}. The monopoles are interacting as Coulomb particles, thus forming a magnetically neutral plasma. In three dimensions the monopoles are pointlike, {\it i.~e.} instanton-like objects. Along the $\kappa_h=\kappa_s=0$ edge the vacuum of the model possesses a mass gap at any non-zero density of monopoles, which is realized at any finite $\beta$. The mass gap is given by the Debye mass of the monopole plasma.

Along the edge $\beta \to \infty$, $\kappa_s=0$ the model describes a superfluid. Indeed, along this edge the gauge field becomes constrained to the trivial vacuum, $\dd \theta=0$. The constraint is resolved as $\theta = \dd \alpha$. Identifying $\alpha + \varphi_h = \varphi$, the model becomes the $3D$ $XY$ spin model
\beqn
  S_{XY} = - \kappa^{XY} \sum\limits_l \cos(\dd\varphi)_l\,,
  \label{eq:XY}
\eeqn
where we have set $\kappa_h \equiv \kappa^{XY}$. The $XY$ model is known to possess a second order transition~\cite{ref:XYphase} at the critical point
\beqn
  \kappa^{XY}_c \approx 0.45420\dots\,,
  \label{eq:XY:critical}
\eeqn
shown by the dot in Figure~\ref{fig:phase:2D:I}.

The edges $\beta=0$ and $\kappa_h \to \infty$ correspond to trivial theories so that they do not posses any phase transitions.

\subsubsection{The $\kappa_s \to \infty$ face: $XY$ spins coupled to $\Z_2$ gauge field}

\begin{figure}[!htb]
  \begin{center}
  \includegraphics[width=6.5cm]{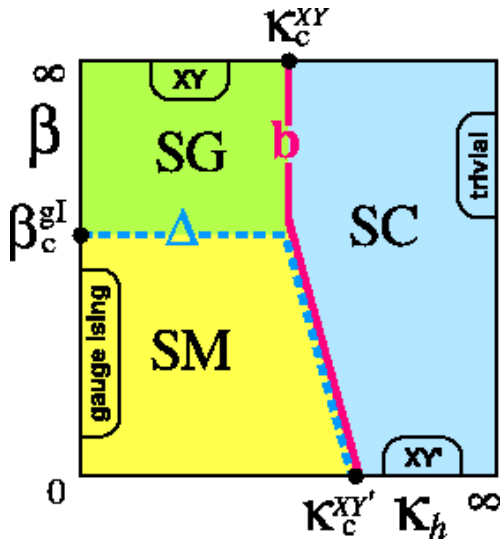}
  \end{center}
  \caption{The phase diagram on the $\kappa_s \to \infty$ face corresponding to the
           model~\eq{eq:XYZ2:action}. The critical points at the edges and the phase transition lines in
	   the interior are discussed in the text.}
  \label{fig:phase:2D:II}
\end{figure}

In the limit $\kappa_s \to \infty$ the coupling of the spinon-pair field and the gauge field is tight. Mathematically, this is expressed in the form of the constraint
\beqn
  2 \theta + \dd \varphi_s = 2 \pi m\,, \quad m \in \Z\,,
  \label{eq:2theta}
\eeqn
which is to be fulfilled at each link. The constraint~\eq{eq:2theta} has the solution
$\theta = (\dd \varphi_s/2 + \pi m)_{2 \pi}$, where $m \in \Z$ is chosen such that
$\theta \in (-\pi,\pi]$. Substituting this solution back into Eq.~\eq{eq:cA2HM:action}, and
introducing the $\Z_2$ gauge field $\sigma_l = {(-1)}^{m_l}$, we get that on the face
$\kappa_s \to \infty$ the model \eq{eq:cA2HM:action} reduces to a $3D$ $XY$-$\Z_2$ model with the action
\beqn
  S_{XY-\Z_2} = -\beta \sum\limits_P \sigma_P -
  \kappa_h \sum\limits_l \sigma_l \cos(\dd\varphi)_l\,.
  \label{eq:XYZ2:action}
\eeqn
This model describes a $XY$-type matter field $\varphi = \varphi_h - \varphi_s/2$ which interacts via the exchange of a $\Z_2$ gauge field $\sigma_l$. The model has a rich phase structure studied numerically in Ref.~\onlinecite{ref:Z2XY}. The phase diagram -- plotted in Figure~\ref{fig:phase:2D:II} -- contains three phases:
\begin{itemize}
  \item[(i)]   the SM phase at small $\beta$/small $\kappa_h$,
  \item[(ii)]  the SG phase at large $\beta$/small $\kappa_h$,
  \item[(iii)] the SC phase at large $\beta$/large $\kappa_h$.
\end{itemize}
As one can see from Figure~\ref{fig:phase:2D:II} the phase structure of the $\kappa_s \to \infty$ face is much richer than that of the $\kappa_s =0$ face.

Let us analyze the edges of this two-dimensional phase diagram. At the $\kappa_h=0$ edge the model is
reduced to the $\Z_2$-gauge (or, the ``gauge Ising'') model~\cite{ref:GaugeIsing} with the action
\beqn
  S_{{\mathrm{gI}}} = - \beta \sum\limits_P \sigma_P \, .
  \label{eq:Z2:gauge}
\eeqn
This model has a string-like topological object as well as vacuum excitations which are sometimes considered as prototypes of, respectively, the chromoelectric string and glueball excitations in the strongly interacting quark-systems in non-Abelian gauge theories~\cite{ref:Caselle}.

The gauge Ising model~\eq{eq:Z2:gauge} is known to possess a second order phase transition of the Ising-type at the critical point
\beqn
  \lim_{\kappa_h \to 0} \beta_c(\kappa_h) = \beta^{\mathrm{gI}}_c \approx 0.7613\dots
  \label{eq:XY:gauge:critical}
\eeqn
shown by the dot at $\kappa_h=0$ edge of Figure~\ref{fig:phase:2D:II}. This transition separates the disordered (small $\beta$) and ordered (large $\beta$) phases discriminated by the presence or absence of the spinon-pair condensate, $\Delta = 0$ and $\Delta \neq 0$, respectively. At the same time, at this axis the holon condensate vanishes due to disorder in the $XY$-variables (to be discussed below). Therefore the small-$\beta$ and large-$\beta$ phase are identified with the SM and SG phases, respectively.

At the $\beta \to \infty$ edge the $\Z_2$ gauge field is suppressed due to the constraint $\sigma_P = 1$, so that the model is reduced to the $XY$ model~\eq{eq:XY} with a coupling $\kappa^{XY} \equiv \kappa_h$. At small values of this coupling, $\kappa_h < \kappa^{XY}_c$, the holon field is disordered by vortices and the holon condensate is absent. At large values $\kappa_h > \kappa^{XY}_c$, the vortices are dilute and
holon condensation takes place. These regimes are separated by the critical coupling~\eq{eq:XY:critical} which is marked by the dot at the $\beta \to \infty$ edge of the phase diagram, Figure~\ref{fig:phase:2D:II}.

The interior of the phase diagram is also nontrivial. The phase transitions, marking the onset of spinon-pair (dotted line) and holon (solid line) condensations, are departing from the $\kappa_h=0$ and $\beta \to \infty$ edges towards the center of the phase diagram. There they meet together forming a single transition line where the holon and spinon-pair condensations occur simultaneously, Figure~\ref{fig:phase:2D:II}. This combined line extends down to lower $\beta$, and finally it meets with the $\beta=0$ edge of the phase diagram. It is described by a (modified) $XY^{\prime}$ model with the action
\beqn
  S_{XY'} = - \sum_l \log \cosh [\kappa_h \cos (\dd \varphi)_l] \, ,
  \label{eq:XY:prime}
\eeqn
which can be derived explicitly from Eq.~\eq{eq:XYZ2:action}. The modified $XY^{\prime}$ model~\eq{eq:XY:prime} possesses a second order phase transition at the critical point~\cite{ref:Z2XY}
\beqn
  \lim_{\beta \to 0}\kappa_{h,c}(\beta) = \kappa^{XY'}_c \approx 1.6\,.
  \label{eq:XY:prime:critical}
\eeqn

The $\kappa_h \to \infty$ edge is trivial, being occupied by the superconducting phase with both the holon and the spinon-pair condensates present.

The model \eq{eq:XYZ2:action} alone was supposed~\cite{ref:fractionalized} to possess an interesting link to the physics of correlated electrons being able to describe certain exotic phases. The topological objects
of this model are called ``visons'' which are fractionally charged excitations. In the language of the
Abelian two-Higgs model the vison coincides with the spinon vortex, while the holon vortex turns into
a $XY$ vortex, $\varphi$ becomes the field of the so-called ``chargon'' particle. The SG phase corresponds to the fractionalized phase where visons are absent and chargons are free particles. In the SM phase the visons are condensed, and chargons are confined. The SC phase corresponds to a superfluid state where both visons and $XY$ vortices are dilute and chargons are free.

\subsection{The $(\beta,\kappa_s)$-faces}

The coupling between the holon field and the gauge field is controlled by the parameter $\kappa_h$.
Below we consider the influence of the spinons on the phase diagram considering the limits of large ($\kappa_h \to \infty$) and small couplings ($\kappa_h \to 0$).

\subsubsection{The $\kappa_h=0$ face: $Q=2$ compact Abelian Higgs model}

On the $\kappa_h = 0$ face the holon condensate vanishes, $b \equiv 0$, and the spinon-pair condensate $\Delta$ is described by the cAHM${}_{Q=2}$ model:
\beqn
  S_{\mathrm{cAHM}_{Q=2}} = -\beta \sum\limits_P \cos\theta_P
  - \kappa_s \sum_l \cos({\mathrm{d}} \varphi_s + 2 \theta)_l \,.
  \label{eq:cAHM2}
\eeqn
The phase diagram of this model in the $(\beta,\kappa_s)$-plane is well known and the behavior of the spinon-pair condensate can be deduced from results of Ref.~\onlinecite{ref:fradkin,Smiseth:2003bk,ref:AHM2}.
Before doing so, let us consider the edges of the phase diagram.

The edges $\kappa_s = 0$ and $\beta = 0$ do not possess any phase transitions because they are described, respectively, by the cQED model~\eq{eq:cQED} and by a trivial model.

The edge $\beta \to \infty$ corresponds to the $XY$ model, since in this limit the constraint
$\dd \theta = 0$ is imposed. The constraint is resolved by setting $\theta = \dd \alpha$, with subsequent identification
\[
  \varphi_s + 2 \alpha - 2 \pi l = \varphi \in (-\pi,\pi]
\]
with $l \in \Z$. Then we obtain the $XY$ action~\eq{eq:XY} by setting $\kappa^{XY} \equiv \kappa_s$.
The $XY$ model describes the superfluid behavior of the spinon-pair condensate with the $XY$ critical point~\eq{eq:XY:critical}, $\kappa^{XY}_c$.

The edge $\kappa_s \to \infty$ corresponds to the gauge Ising model~\eq{eq:Z2:gauge} because in this limit the constraint
\[
  {\mathrm{d}} \varphi_s + 2 \theta = 2 \pi m\,, \quad m \in \Z\,,
\]
is automatically imposed. The constraint is resolved as
\[
  \theta = \pi m - {\mathrm{d}} \varphi_s/2 + 2 \pi n\,, \quad n \in \Z\,,
\]
leading subsequently to
\[
  \dd \theta = \pi \dd m\,, \quad \cos \dd \theta = \cos \pi \dd m \equiv \sigma_P\,,
\]
where $\sigma_l = {(-1)}^{m_l}$ is the $\Z_2$ gauge field attached to the link $l$.
Then Eq.~\eq{eq:cAHM2} reduces to Eq.~\eq{eq:Z2:gauge}, and the edge $\kappa_s \to \infty$ possesses
a critical point at $\beta_c = \beta^{\mathrm{gI}}_c$ which marks the second order phase transition
of Ising type.

\begin{figure}[!htb]
  \begin{center}
  \includegraphics[width=7.0cm]{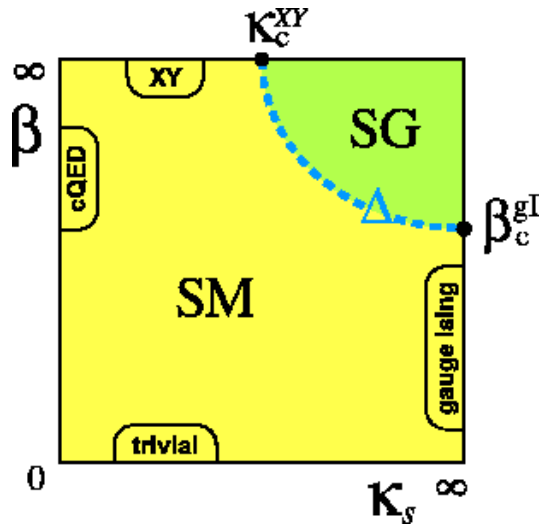}
  \end{center}
  \caption{The phase diagram of the compact Abelian $Q=2$ Higgs model~\eq{eq:cAHM2}
           on the $\kappa_h=0$ face.}
  \label{fig:phase:2D:III}
\end{figure}

The critical points at the $\beta \to \infty$ and $\kappa_s \to \infty$ edges are connected by a
second order phase transition line as shown by the dotted line in Figure~\ref{fig:phase:2D:III}.
The SM phase is located at small $\beta$ and/or small $\kappa_s$, while the SG phase is residing
in the large $\beta$/large $\kappa_s$ corner. The transition line corresponds to the onset of
spinon-pair condensation $\Delta$. In the whole diagram the holon condensate is absent, $b=0$.
In addition also the limiting models ($XY$, cQED, gauge Ising and trivial cases) together with the
critical points $\kappa^{XY}_c$, Eq.~\eq{eq:XY:critical}, and $\beta^{\mathrm{gI}}_c$, Eq.~\eq{eq:XY:gauge:critical}, are indicated.

\subsubsection{The $\kappa_h \to \infty$ face: $XY$ model}

\begin{figure}[!htb]
  \begin{center}
  \includegraphics[width=6.0cm]{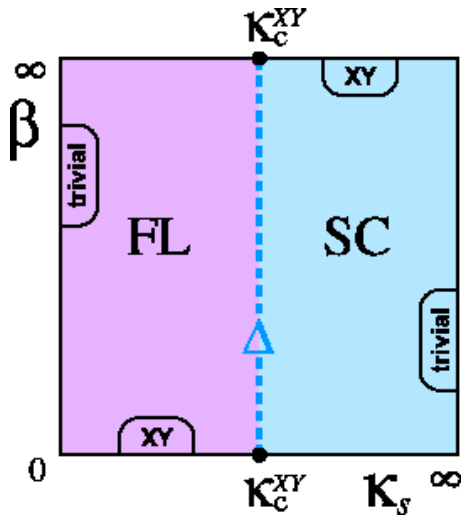}
  \end{center}
  \caption{The phase diagram of the $XY$ model~\eq{eq:XY} together with the limiting models at the edges
           on the $\kappa_h \to \infty$ face.}
  \label{fig:phase:2D:IV}
\end{figure}

On the $\kappa_h \to \infty$ face the model \eq{eq:cA2HM:action} reduces to the $XY$ model~\eq{eq:XY}
with a hopping parameter $\kappa \equiv \kappa_s$ and $\varphi= - 2 \varphi_h +\varphi_s$. The model controls the superfluid behavior of the spinon-pair condensate $\Delta$. Due to the constraint
\[
  \theta = \dd \varphi_h + 2 \pi l\,, \quad l \in \Z\,,
\]
the holon vortices are suppressed and therefore $b \neq 0$ in the whole $(\beta$,$\kappa_s)$-plane.
The phase diagram -- shown in Figure~\ref{fig:phase:2D:IV} --
is divided by a second order $XY$-like transition line parallel to the $\beta$ axis at
$\kappa_{s,c}(\beta) = \kappa^{XY}_c$ shown as dotted line. This line separates the FL phase (with condensed spinon vortices and $\Delta=0$) at $\kappa_s < \kappa^{XY}_c$ from the SC phase (with suppressed spinon vortices and $\Delta\ne0$) at $\kappa_s > \kappa^{XY}_c$. In the whole diagram the holon field is condensed, $b \neq 0$.

\subsection{The $(\kappa_h,\kappa_s)$-faces}

Now we consider the effect of choosing the extreme limits of very strong and very weak gauge couplings, $\beta=0$ and $\beta \to \infty$, respectively.

\subsubsection{The $\beta=0$ face: an ultralocal two-Higgs system}

On the $\beta=0$ face we obtain a two-Higgs system interacting ultra-locally via a non-propagating gauge field. The one-link action of the model is given by
\beqn
  e^{- S_l[\varphi_h,\varphi_s]} = \! \int \!\dD \theta \exp\Bigl\{
  \sum_{k=h,s} \kappa_k \cos({\mathrm{d}} \varphi_k + q_k \theta)_l\Bigr\}\,.
  \label{eq:2Higgs:ultralocal:1}
\eeqn
In order to get this part of the weight for the limiting model we put $\beta=0$ in Eq.~\eq{eq:cA2HM:action} and keep the integration over the gauge field $\theta$. We expand the two factors of the exponentiated one-link action in Eq.~\eq{eq:2Higgs:ultralocal:1} in a Fourier series,
\beqn
  e^{\kappa_k \cos({\mathrm{d}} \varphi_k + q_k \theta)_l} = \sum_{n_k \in \Z} I_{n_k}(\kappa_k)\,
  e^{i \, [({\mathrm{d}} \varphi_k + q_k \theta)_l] n_k}\,,
  \label{eq:2Higgs:ultralocal:2}
\eeqn
for $q_h=1$ and $q_s=2$. Here $I_n(\kappa)$ is the modified Bessel function of $n$-th order.
Substituting Eq.~\eq{eq:2Higgs:ultralocal:2} in Eq.~\eq{eq:2Higgs:ultralocal:1} and performing the integration over $\theta$ we obtain the constraint $n_h + 2 n_s = 0$.
Setting $n_s = - n_h/2 = n$, we get (up to an inessential factor in front of the sum)
\beqn
  e^{- S_l[\varphi_h,\varphi_s]} = \sum_{n \in \Z} I_{2n}(\kappa_h)\, I_n(\kappa_s)\,
  e^{i \, {\mathrm{d}} (\varphi_s - 2\varphi_h) n}\,,
  \label{eq:2Higgs:ultralocal}
\eeqn
where we used the property $I_n \equiv I_{-n}$. Note that the quantities $n$, $\dd \varphi_h$,  $\dd \varphi_s$ in Eq.~\eq{eq:2Higgs:ultralocal} are defined at the same link $l$.

\begin{figure}[!htb]
  \begin{center}
  \includegraphics[width=7.0cm]{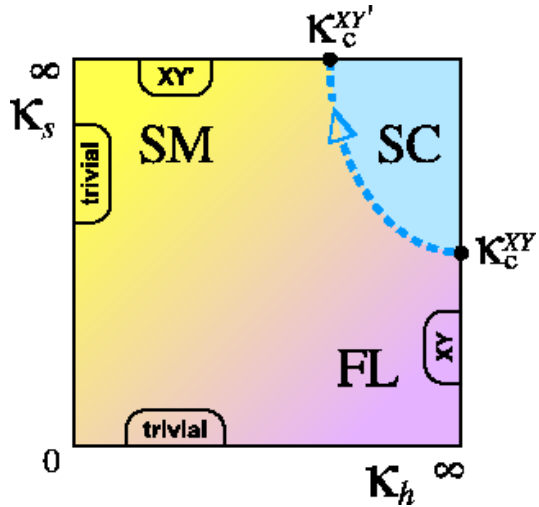}
  \end{center}
  \caption{The phase diagram of the ultralocal two-Higgs system~\eq{eq:2Higgs:ultralocal} on the
           $\beta=0$ face.}
  \label{fig:phase:2D:V}
\end{figure}

In the $(\kappa_h$,$\kappa_s)$-plane one has two phases: the SC phase with non-zero condensates $\Delta$ and $b$ in the large $\kappa_h$/large $\kappa_s$ corner and a SM-FL phase in the remaining part of the phase diagram. The phase diagram is shown in Figure~\ref{fig:phase:2D:V}. The SC and SM-FL phases are separated by a second order $XY$-type phase transition (indicated by a dotted line) which starts at $\kappa_{h,c} = \kappa^{XY'}_c$, Eq.~\eq{eq:XY:prime:critical}, at the $\kappa_s \to \infty$ edge and
ends at $\kappa_{s,c} = \kappa^{XY}_c$, Eq.~\eq{eq:XY:prime}, for $\kappa_h \to \infty$. At these two
edges the two-Higgs model~\eq{eq:2Higgs:ultralocal} is reduced to a modified~\eq{eq:XY:prime} and a usual~\eq{eq:XY} $XY$ model, respectively. The SM-FL phase appears actually as the FL phase ($b \neq 0$) at large $\kappa_h$, and the SM phase ($b \to 0$) is realized at large $\kappa_s$,
(the structure of the SM-FL phase is plotted in Figure~\ref{fig:phase:2D:I}).

\subsubsection{The $\beta \to \infty$ face: two decoupled $XY$ models}

\begin{figure}[!htb]
  \begin{center}
  \includegraphics[width=7.0cm]{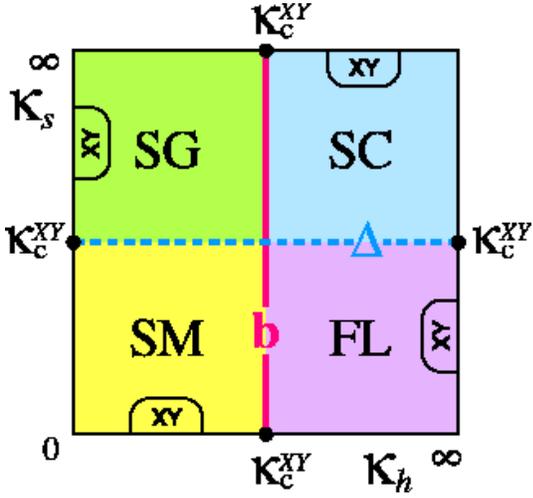}
  \end{center}
  \caption{The phase diagram on the $\beta \to \infty$ face of the two decoupled $XY$ models~\eq{eq:2XY}.}
  \label{fig:phase:2D:VI}
\end{figure}

Finally, on the $\beta \to \infty$ face the system~\eq{eq:cA2HM:action} reduces to two decoupled $XY$
models describing the holon and spinon-pair superfluids. This fact is readily seen from the partition function~\eq{eq:cA2HM:action}: large $\beta$ imposes the constraint $\dd \theta = 0$ which is resolved,
as usual, by $\theta = \dd \alpha + 2 \pi n$, $n \in \Z$. Substituting this solution back to~\eq{eq:cA2HM:action} and performing the redefinitions of the phases,
$\varphi_h + \alpha \to \varphi_h$ and $\varphi_s + 2 \alpha \to \varphi_s$, one gets
\beqn
  S_{2XY} = - \kappa_h \sum_l \cos(\dd\varphi_h)_l - \kappa_s \sum_l \cos(\dd\varphi_s)_l \, .
  \label{eq:2XY}
\eeqn
The phase diagram in the $(\kappa_h,\kappa_s)$-plane includes all discussed phases (SM, SG, SC and FL) as shown in Figure~\ref{fig:phase:2D:VI}.
The phases are separated by the straight solid (condensation of the holon $b$) and dashed
(condensation of the spinon-pair $\Delta$) phase transition lines at $\kappa_{h,c} =  \kappa^{XY}_{c}$
and $\kappa_{s,c} = \kappa^{XY}_{c}$, Eq.~\eq{eq:XY:critical}, respectively.

\subsection{The interior of the $3D$ phase diagram}

\begin{figure}[!htb]
  \includegraphics[width=8.0cm]{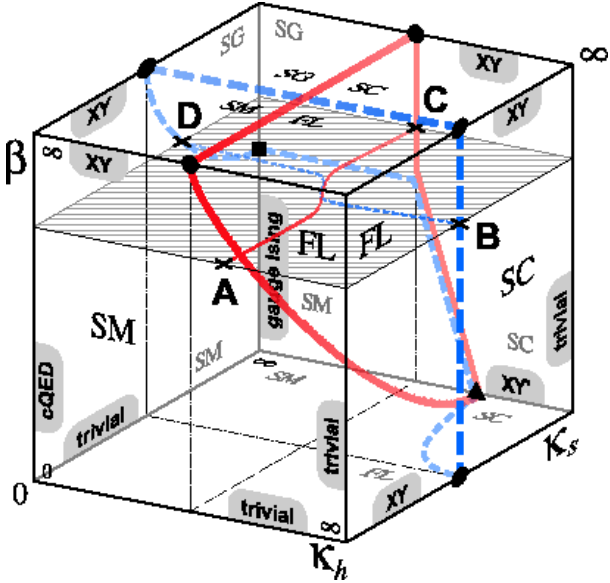}
  \caption{The qualitative $3D$ phase diagram of the cA2HM (reproduced from
           Ref.~\onlinecite{ref:PRB2006}).}
  \label{fig:phase:3D}
\end{figure}

Knowing the limiting cases shown in Figures~\ref{fig:phase:2D:I}-\ref{fig:phase:2D:VI} allows us
(following Ref.~\onlinecite{ref:PRB2006}) to reconstruct the interior of the three-dimensional phase
diagram as shown schematically in Figure~\ref{fig:phase:3D}. In other words, the six faces of the $3D$ cube
Figure~\ref{fig:phase:3D} correspond to the $2D$ diagrams plotted in Figures~\ref{fig:phase:2D:I}-\ref{fig:phase:2D:VI}.

\begin{figure}[!htb]
  \begin{center}
  \begin{tabular}{cc}
    \includegraphics[width=4.2cm]{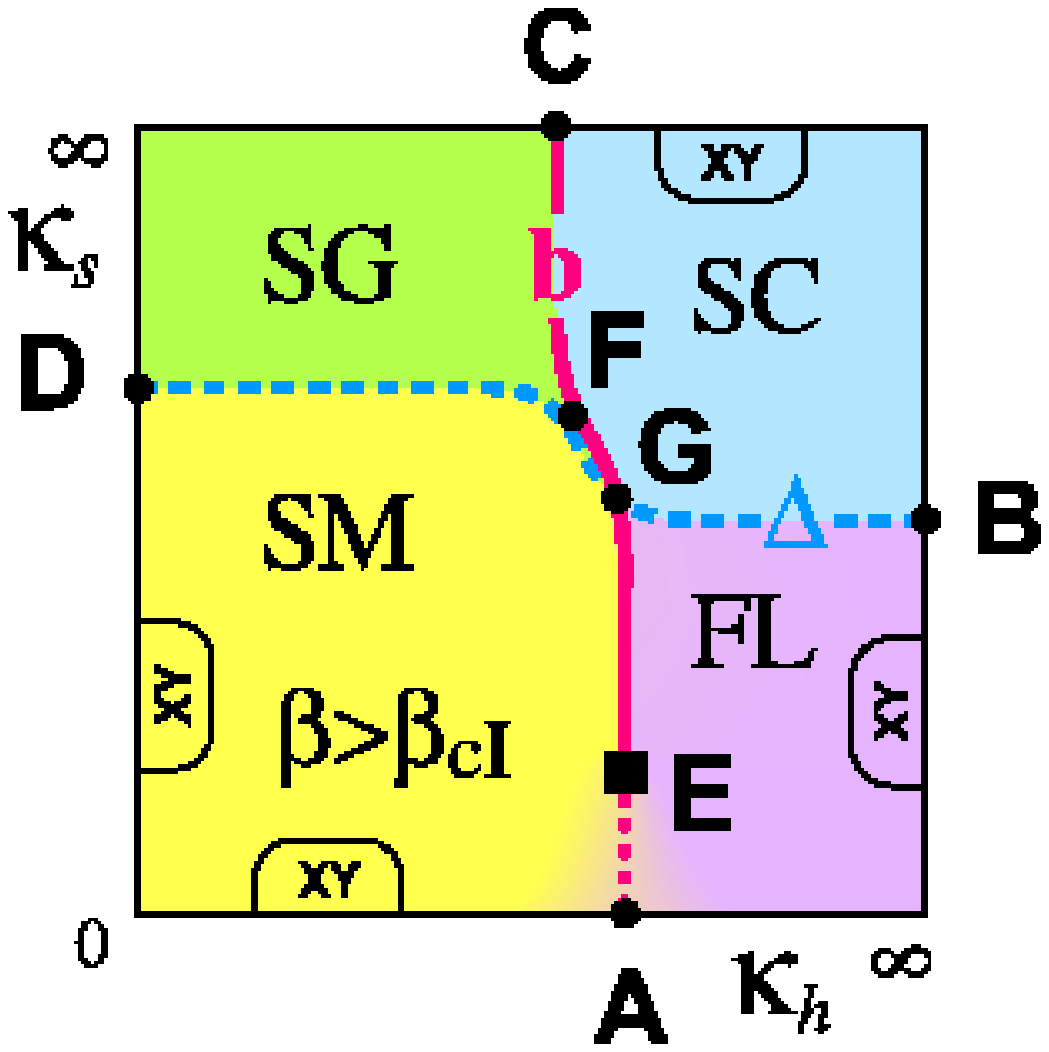} &
    \includegraphics[width=4.2cm]{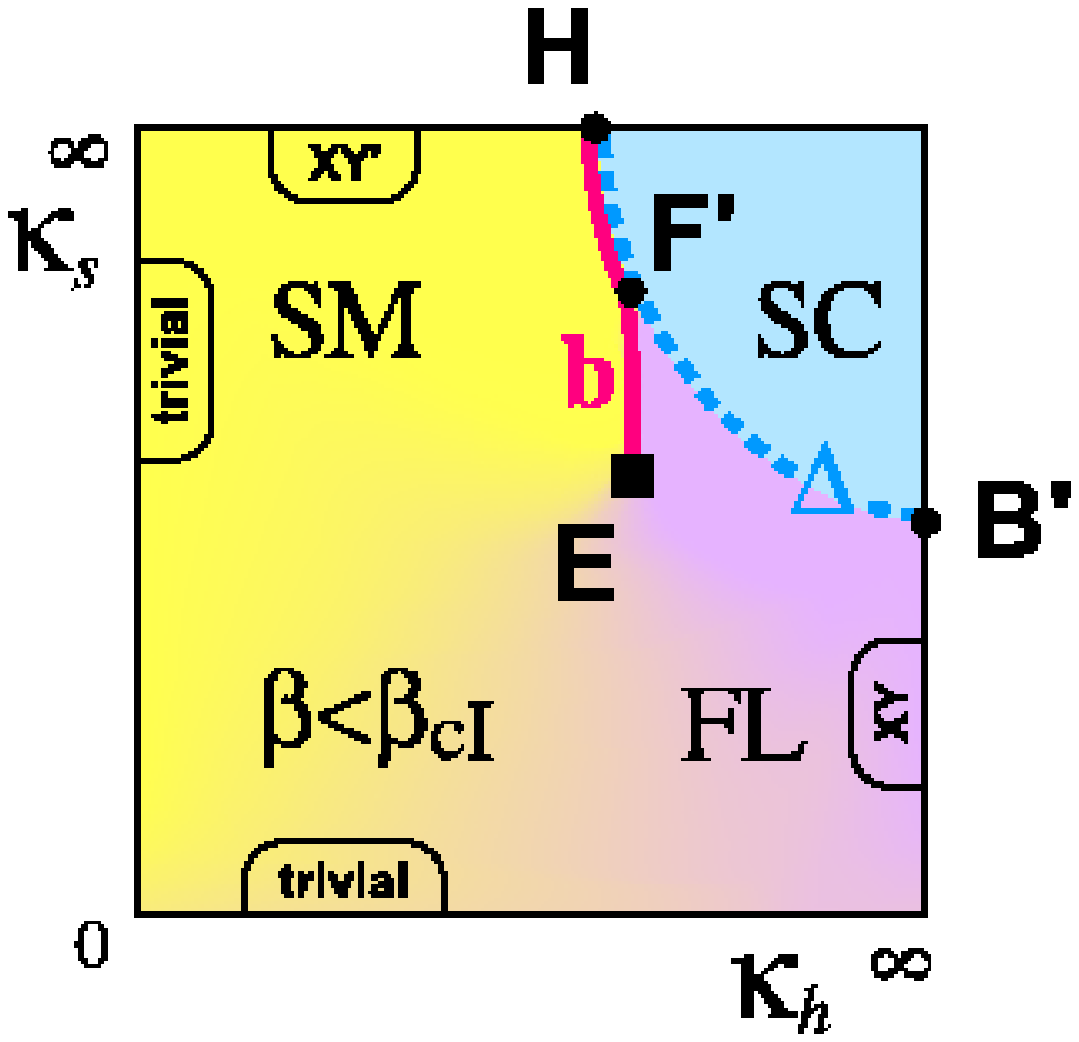} \\[1mm]
    (a)  &  (b) \\[3mm]
    \includegraphics[width=4.2cm]{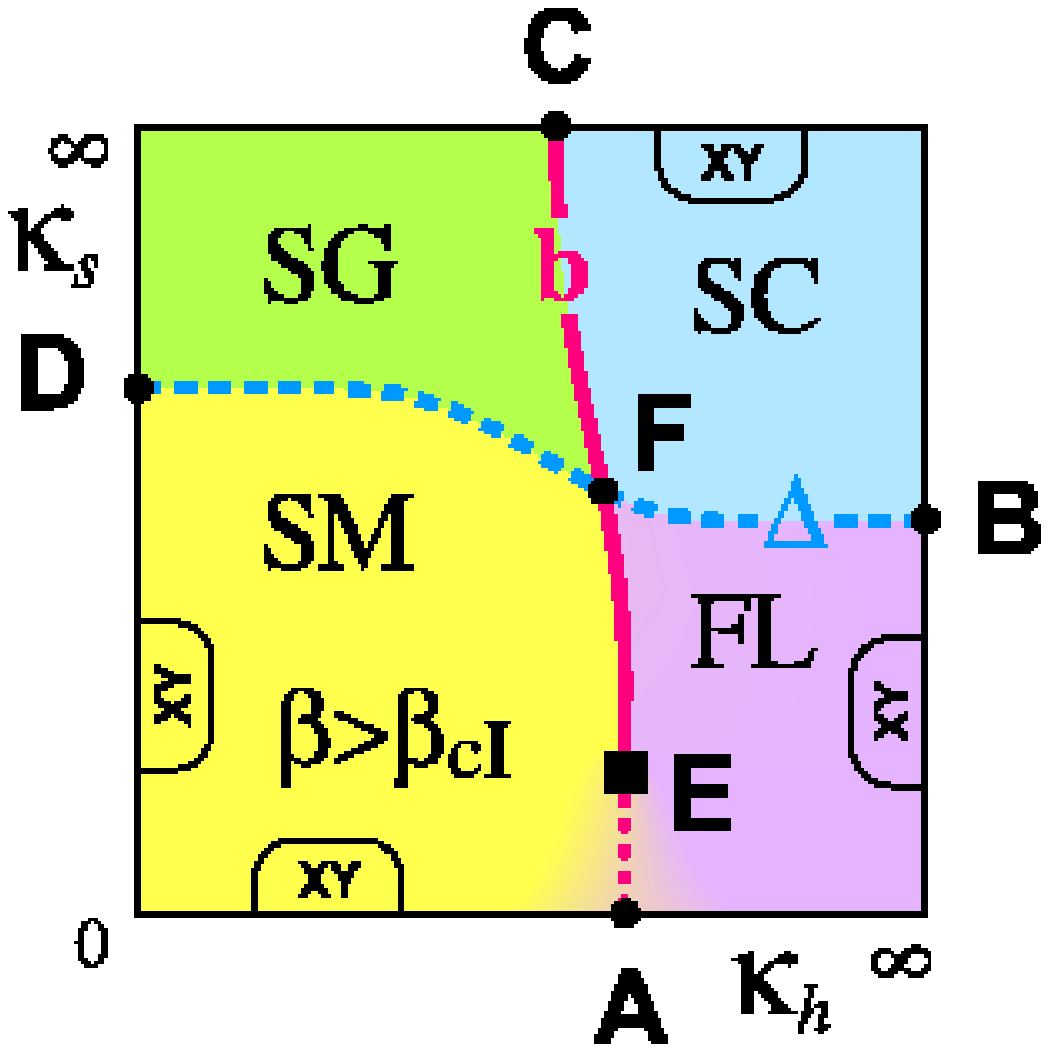} &
    \includegraphics[width=4.2cm]{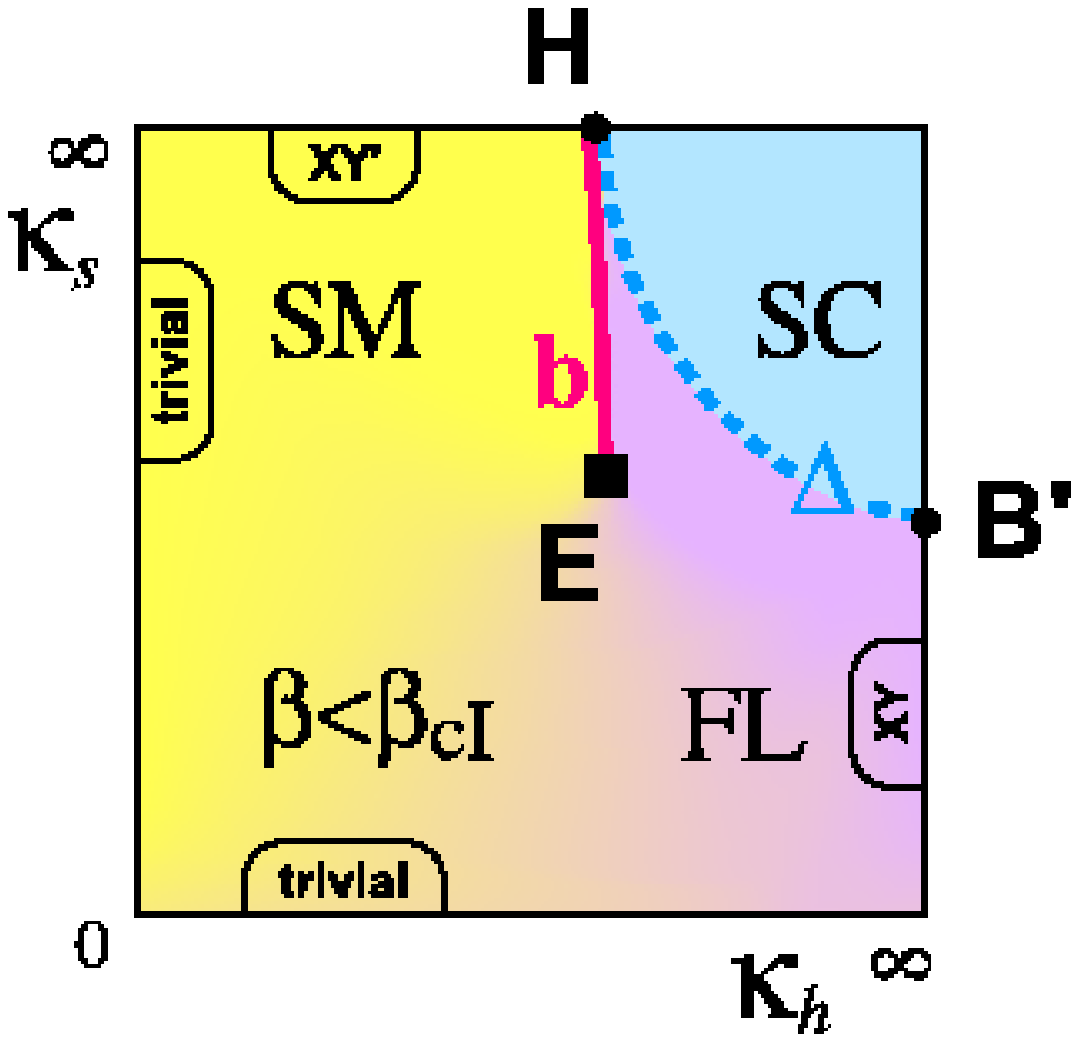} \\[1mm]
    (c)  &  (d)
  \end{tabular}
  \end{center}
  \caption{The $2D$ section of the $3D$ phase cube at a fixed gauge coupling $\beta$:
           (a) $\beta_c^{\mathrm{gI}} < \beta < \infty$ and
           (b) $0 < \beta < \beta_c^{\mathrm{gI}}$.
            The meaning of the points A,B,... is explained in the text. In both Figures in the middle
	    of the diagram the transition lines of $b$- and $\Delta$-condensation
            join along a finite segment. Figures (c) and (d) represent the respective alternative scenario
            with the transition lines intersecting instead of piecewise merging together.}
  \label{fig:phase:2D:section1}
\end{figure}

The shaded section in Figure~\ref{fig:phase:3D} represents the phase diagram in the $(\kappa_h,\kappa_s)$-plane at fixed finite gauge coupling $\beta$ in the region
$\beta_c^{\mathrm{gI}} < \beta < \infty$. {\it A priori} there are two possible views to expect of the internal section of the phase diagram. One of the options is plotted schematically in Figure~\ref{fig:phase:2D:section1}~(a). The holon and spinon-pair condensation lines are getting slightly curved with respect to the limiting ($\beta \to \infty$) case shown in Figure~\ref{fig:phase:2D:VI}:
the solid (dashed) line, which marks the condensation of the holon $b$ (spinon-pair $\Delta$),
gets generally shifted towards larger values of $\kappa_h$ ($\kappa_s$). The holon condensation line starts at large values of $\kappa_s$ at the point
\[
  \kappa_{h,c}(\beta) > \kappa_{h,c}(\infty) \equiv \kappa^{XY}_c\,.
\]
This point is marked as the point C in Figures~\ref{fig:phase:3D} and \ref{fig:phase:2D:section1}~(a).
With $\kappa_s$ becoming smaller the $b$-condensation line meets the $\Delta$-condensation line at the point F, and they continue together till the point G, at which these transition split again. Then the $b$-condensation line continues alone and eventually stops at an endpoint E in the interior of the diagram.
A projection of this line to the edge $\kappa_s=0$ (marked in Figure~\ref{fig:phase:2D:section1}~(a)
by the dotted line  E-A) might eventually be visible as a percolation transition.
The $\Delta$-condensation transition, denoted by the dashed line D-F-G-B, does not have an endpoint.

The two-dimensional section qualitatively changes as $\beta$ gets smaller than the critical coupling $\beta_c^{\mathrm{gI}}$ of the gauge Ising model. An example of such a section is shown in Figure~\ref{fig:phase:2D:section1}~(b). At sufficiently strong gauge coupling,
$\beta < \beta^{\mathrm{gI}}_c$, the points C, D and F merge together into the point denoted now as H and
the SG phase pocket disappears. The $b$- and $\Delta$-condensation lines begin to run together from the point H till the point F$^{\prime}$ (formerly point G in Figure~\ref{fig:phase:2D:section1}~(a))
in the interior of the phase diagram, where the lines split again: the $\Delta$-condensation
goes to the point B$^{\prime}$, while the $b$-condensation line runs towards the $\kappa_s=0$
edge but ends at a new endpoint E. As $\beta$ decreases further, Figure~\ref{fig:phase:2D:section1}~(b)
gradually evolves into Figure~\ref{fig:phase:2D:V}: at $\beta=0$ the endpoint
E becomes finally the point H. As a result, the $b$-condensation transition disappears completely,
as it is plotted in Figure~\ref{fig:phase:2D:V}.

The alternative to the above scenario -- which cannot be excluded by analytical means -- is represented in Figures~\ref{fig:phase:2D:section1}~(c) and (d). The difference is in the mutual behavior of the $b$- and $\Delta$-condensation lines in the interior of the phase diagram. In Figure~\ref{fig:phase:2D:section1}~(a) these transitions piecewise join along a common segment F-G. In Figure~\ref{fig:phase:2D:section1}~(b)
the common segment is H-F$^{\prime}$. The alternative of Figure~\ref{fig:phase:2D:section1}~(a)
is plotted in Figure~\ref{fig:phase:2D:section1}~(c): instead of merging along the F-G segment the transition lines intersect in the point F. The alternative of Figure~\ref{fig:phase:2D:section1}~(b)
is plotted in Figure~\ref{fig:phase:2D:section1}~(d): instead of having the common segment H-F$^{\prime}$ (in Figure~(b)) the $b$- and $\Delta$-transition lines have only a common starting point H and separate immediately.

To discriminate between these scenarios numerical simulations have to be used. This is presented in the next section.

\section{Numerical results}
\label{sec:numerical}

\subsection{Observables}

In order to clarify the structure of the phase diagram we have performed a numerical study of various gauge-invariant quantities. One potentially sensitive quantity is the action of the model, or any  (gauge-invariant) part of it. This fact is easy to understand since the action governs the dynamics of
the whole model. Using a familiar notation, the action of our lattice model is given as
\beqn
  S&=&-\beta \, S_P - \kappa_h \, S_h - \kappa_s \, S_s
\eeqn
with
\beqn
  &&S_P = \sum_{x,\mu<\nu}
  \cos(\theta_{x,\mu} + \theta_{x+\hat\mu,\nu}-\theta_{x+\hat\nu,\mu}-\theta_{x.\nu})  \,,
  \nonumber
  \\
  &&S_h = \sum_{x,\mu} \cos(\varphi_h(x+\hat\mu) - \varphi_h(x) + \theta_{x,\mu})  \,,
  \\
  &&S_s = \sum_{x,\mu}  \cos(\varphi_s(x+\hat\mu) - \varphi_s(x) + 2 \, \theta_{x,\mu})  \,.
  \nonumber
\eeqn
Here $x$ denotes the sites of the $3D$ lattice, nearest neighbors are separated by a lattice spacing $a$,
and $\hat \mu$ is the shift vector in the $\mu$ direction. The compact lattice gauge field angles $\theta_{\mu}(x) \in (-\pi,\pi]$ live on the links (bonds) between sites  $x$ and $x+\hat\mu$,
the holon fields $\varphi_h(x)$ (phase angles of a singly charged scalar field) and the spinon-pair fields $\varphi_s(x)$ (phase angles of a doubly charged scalar field) are defined on the sites. The coupling $\beta=1/g^2$ is the inverse gauge coupling squared, and $\kappa_h$ and $\kappa_s$ express the coupling between Higgs and gauge fields, respectively. In the following $\kappa_h$ and $\kappa_s$ are called ``hopping parameters''. To study this model by means of Monte Carlo we use standard Metropolis updates for all three kinds of fields.

In order to characterize the different phases of the model we consider the following ``thermodynamical'' expectation values (related to the derivatives of the logarithm of the partition function with respect to the couplings):
\beqn
  \langle E_P \rangle &=& \langle \frac{1}{N_P} S_P \rangle
  \nonumber \\
  \langle E_h\rangle  &=& \langle \frac{1}{N_L} S_h \rangle \,, \\
  \langle E_s\rangle  &=& \langle \frac{1}{N_L} S_s \rangle \,,
  \nonumber
  \label{eq:thdvariables}
\eeqn
called plaquette, holon link and spinon link expectation values. Here $N_{P/L/S}$ is the number of plaquettes/links/sites on the finite lattice ($N_P = N_L$ in three dimensions), $\langle \cdots \rangle$ denotes the ensemble average over configurations. In addition, the susceptibilities of these quantities (related to the second derivatives of the logarithm of the partition function with respect to the couplings)
have been considered:
\beqn
  \chi_{{}_{E_P}} & = & N_P \Bigl(\langle E_P^2 \rangle - \langle E_P \rangle^2 \Bigr) \,,  \\
  \chi_{{}_{E_{h,s}}} & = & N_L \Bigl(\langle E_{h,s}^2 \rangle - \langle E_{h,s} \rangle^2 \Bigr)
  \,. \nonumber
\eeqn

The simplest characteristics of a topological defect is its density. Using the notations of Section~\ref{sec:topdefects} the monopole and the vortex densities are defined as, respectively,
\beqn
  \rho_{\mathrm{mon}} = \frac{1}{N_S} \sum_{\dual c_3} |\dual j| \,,\qquad
  \rho_{{h,s}} = \frac{1}{N_L} \sum_{\dual c_2} |\dual \sigma_{h,s}| \,.
\eeqn
Here $\dual c_3$ are the sites of the dual lattice dual to the cubes of the original lattice, and
$\dual c_2$ are the links of the dual lattice dual to the plaquettes of the original lattice. The monopole charge is defined in the standard way,
\beqn
  j = \frac{1}{2 \pi} \dd {[\dd \theta]}_{2\pi} \,,
\eeqn
where
${[\cdots]}_{2\pi}/(2\pi)$ denotes the integer part modulo $2 \pi$. Note that the plaquette angle
$\dd \theta$ lies in the range $-\pi$ to $\pi$ plus or minus integer multiples of $2 \pi$.
Thus, effectively using the Gauss's law, one considers a forward cube from a lattice position $x$ built up by six plaquettes, defines the integer-valued so-called oriented Dirac strings $n$ passing through these plaquettes and sums over these integers assigned to the outward pointing oriented strings.
This sum defines the monopole number in the cube corresponding to a point of the dual lattice. Following Ref.~\onlinecite{ANO_definition}, the holon and spinon vortex currents are defined as
\beqn
  \sigma_k = \frac{1}{2 \pi} (\dd {[\dd \varphi +  q_k \, \theta]}_{2\pi} - q_k
  {[\dd \theta]}_{2\pi}) \,, \quad k = h, s\,.
\eeqn
These integer-valued oriented currents pierce a given plaquette corresponding to a link of the dual lattice. Together with $\rho_{\mathrm{mon}} $ and $\rho_{h,s}$ we measure also the corresponding susceptibilities
\beqn
  \chi_{{}_{\rho_{\mathrm{mon}}}} & = &
  N_S \Bigl( \langle \rho_{\mathrm{mon}}^2 \rangle
           - \langle \rho_{\mathrm{mon}} \rangle^2 \Bigr) \,,  \\
  \chi_{{}_{\rho_{h,s}}} & = &
  N_L \Bigl(\langle \rho_{h,s}^2 \rangle
          - \langle \rho_{h,s} \rangle^2 \Bigr)
  \,. \nonumber
\eeqn

In this Section we report on Monte Carlo studies of the phase transitions within two-dimensional cross-sections of the whole phase diagram characterized by certain fixed values of the (inverse) gauge coupling $\beta$. These planes are parameterized by the two hopping parameters $\kappa_h$ and $\kappa_s$.

It is important to note that we have to distinguish between the cross-sections lying above and below
the critical value $\beta_c^{\mathrm {gI}} \approx 0.7613$, corresponding to the phase transition
of the pure gauge Ising model. As we have discussed above, the qualitative structure of the phase diagram is different above and below this phase transition, obtained by moving the shaded square in Figure~\ref{fig:phase:3D} up and down. The SG phase (the phase with zero holon condensate and non-zero spinon-pair condensate) is absent in the $\beta < \beta_c^{\mathrm {gI}}$ (strong gauge coupling) region contrary to the $\beta > \beta_c^{\mathrm {gI}}$ (weak gauge coupling) part of the phase diagram.

In our previous work~\cite{ref:PRB2006} we restricted ourselves to the larger $\beta$ region and presented results mainly for $\beta=1.0$. In order to obtain the gross features of the different phases,
we perform the initial studies on $16^3$ lattices, focussing at different fixed values of $\beta$.
Based on the expectations as described in the previous Section~\ref{sec:phasediagram}, we have used a
dense grid of points spanning the ($\kappa_h$,$\kappa_s$) hopping parameter plane over an range
where we expect a nontrivial behavior of the model.

\subsection{Phase structure at weak gauge coupling}

\begin{figure}[!htb]
  \begin{center}
  \includegraphics[width=7.5cm]{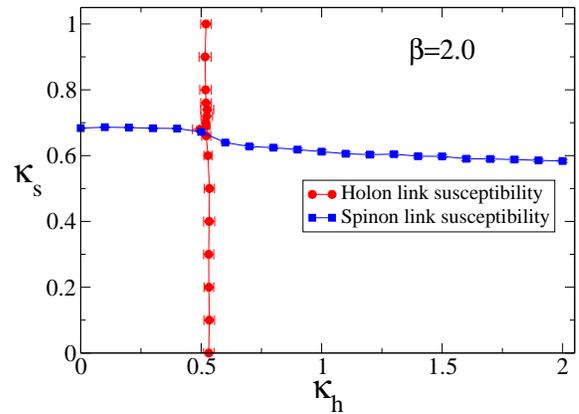}
  \end{center}
  \caption{The two-dimensional cross-section of the three-dimensional phase diagram
           (Figure~\ref{fig:phase:3D}) at $\beta=2.0$, extracted from a $16^3$ lattice using
	   susceptibilities of thermodynamical observables.}
  \label{fig:beta20}
\end{figure}

\begin{figure}[!htb]
  \begin{center}
  \includegraphics[width=7.5cm]{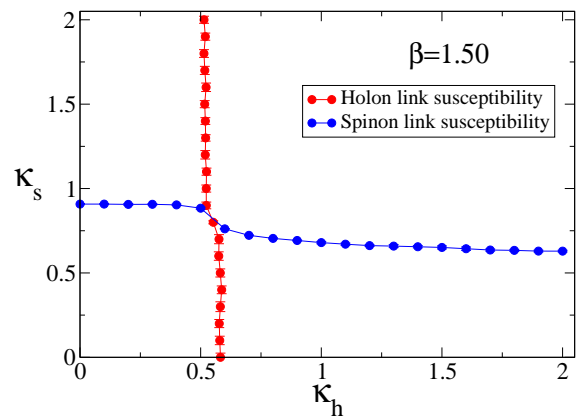}
  \end{center}
  \caption{The same as Figure~\ref{fig:beta20}, but for $\beta=1.5$.}
  \label{fig:beta15}
\end{figure}

\begin{figure}[!htb]
  \begin{center}
  \includegraphics[width=7.5cm]{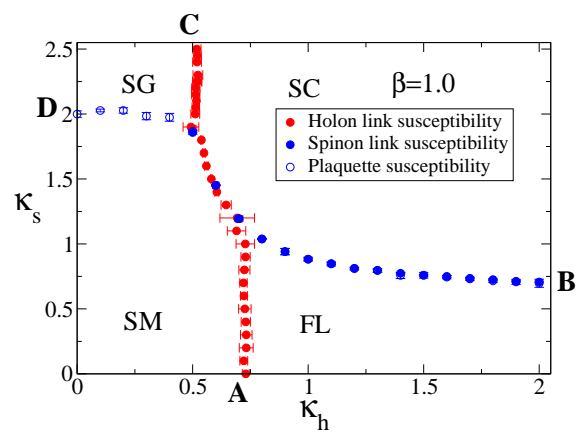}
  \end{center}
  \caption{The same as Figure~\ref{fig:beta20}, but for $\beta=1.0$.
           In addition, the four phases and positions of the points
            A, B, C and D of Figure~\ref{fig:phase:3D} are explicitly indicated.}
  \label{fig:beta10}
\end{figure}

According to our analysis presented above, in the weak coupling region, $\beta>\beta_c^{\mathrm{gI}}$,
the phase diagram contains all four phases. The summary of our results on the phase structure of the model is presented in Figures~\ref{fig:beta20}-\ref{fig:beta10}. These Figures show the two-dimensional
cross-sections -- explored, respectively, at $\beta=2.0$, $\beta=1.5$ and $\beta=1.0$
values of the gauge coupling -- of the full three-dimensional phase diagram.

The non-trivial signals in the measured susceptibilities indicate the existence of transitions between all four different phases known already from our discussion of the weak coupling limit $\beta \to \infty$, presented in Section~\ref{sec:phasediagram}. One can clearly observe that at the largest measured $\beta$ the four different phases meet (within our numerical resolution) in a single crossing point.
The numerically observed picture is very similar to the expected behavior in the limit
$\beta \to \infty$ as shown in Figure~\ref{fig:phase:2D:VI}. The horizontal (blue) and vertical (red) phase transitions are of second order, and both belong to the $XY$ universality class. Note that here we do not discriminate between the ordinary and ``inverted''~\cite{ref:XY:inverted} universality classes of the $XY$ transitions as they are related by a turnover of the coupling axis, $\kappa \to 1/\kappa$, in the course of a duality transformation.

With decreasing $\beta$, the gauge coupling becomes stronger and the phase picture changes (see Fig.~\ref{fig:beta10}). Consider first the horizontal transition line which marks the condensation of spinons. At vanishing holon coupling, $\kappa_h=0$, the phases are separated at a certain ``lower border'' value (point D) of the critical spinon coupling, $\kappa_s^{\mathrm{crit}}(\kappa_h\to 0,\beta)$, which turns out to be a rapid function of the gauge coupling $\beta$. Indeed, as the coupling $\beta$
decreases, the critical spinon coupling $\kappa_s^{\mathrm{crit}}$ substantially increases.
On the other hand, at large values of the holon coupling $\kappa_h \to \infty$, the ``upper border''
(point B) of the critical spinon coupling $\kappa_s^{\mathrm{crit}}(\kappa_h\to \infty, \beta)$
is practically insensitive to the variation of $\beta$.

The features of evolution of the vertical transition line -- which corresponds to the condensation
of the holon pairs -- is similar to the evolution of the horizontal line. The lower endpoint
of the line (point A) evolves moderately while the upper endpoint (point C) practically does
not move at all. As we have already pointed out in Ref \onlinecite{ref:PRB2006}, the indicated line for small $\kappa_s$ at $\beta=1.0$ near the point A does no longer belong to a real phase transition, but characterizes a percolation transition from the SM to the SL phase. This is in agreement with our findings at lower $\beta$'s discussed below.

It is very interesting to find, what happens with these two transition lines in the interior of
the phase diagram. As it is already clear from Figure~\ref{fig:beta15}, the two transition lines
do not simply cross each other in an isolated point. As the inverse gauge coupling $\beta$ gets lower,
in the middle of the phase diagram the transition lines become closer and closer to each other
in a particular interval of the ($\kappa_s$,$\kappa_h$) coupling space. The further decrease of the coupling $\beta$ leads to a qualitative change of the picture in the interior, as it is indicated in
Figure~\ref{fig:beta10}: at $\beta=1.0$ the transition lines piecewise join into a single line in a
certain $(\kappa_h,\kappa_s)$ region. In this region we observe non-trivial signals for a first order transition in all measured quantities to be discussed in the next subsection.

\begin{figure}[!htb]
  \begin{tabular}{cc}
  \multicolumn{2}{c}{\includegraphics[width=4.2cm]{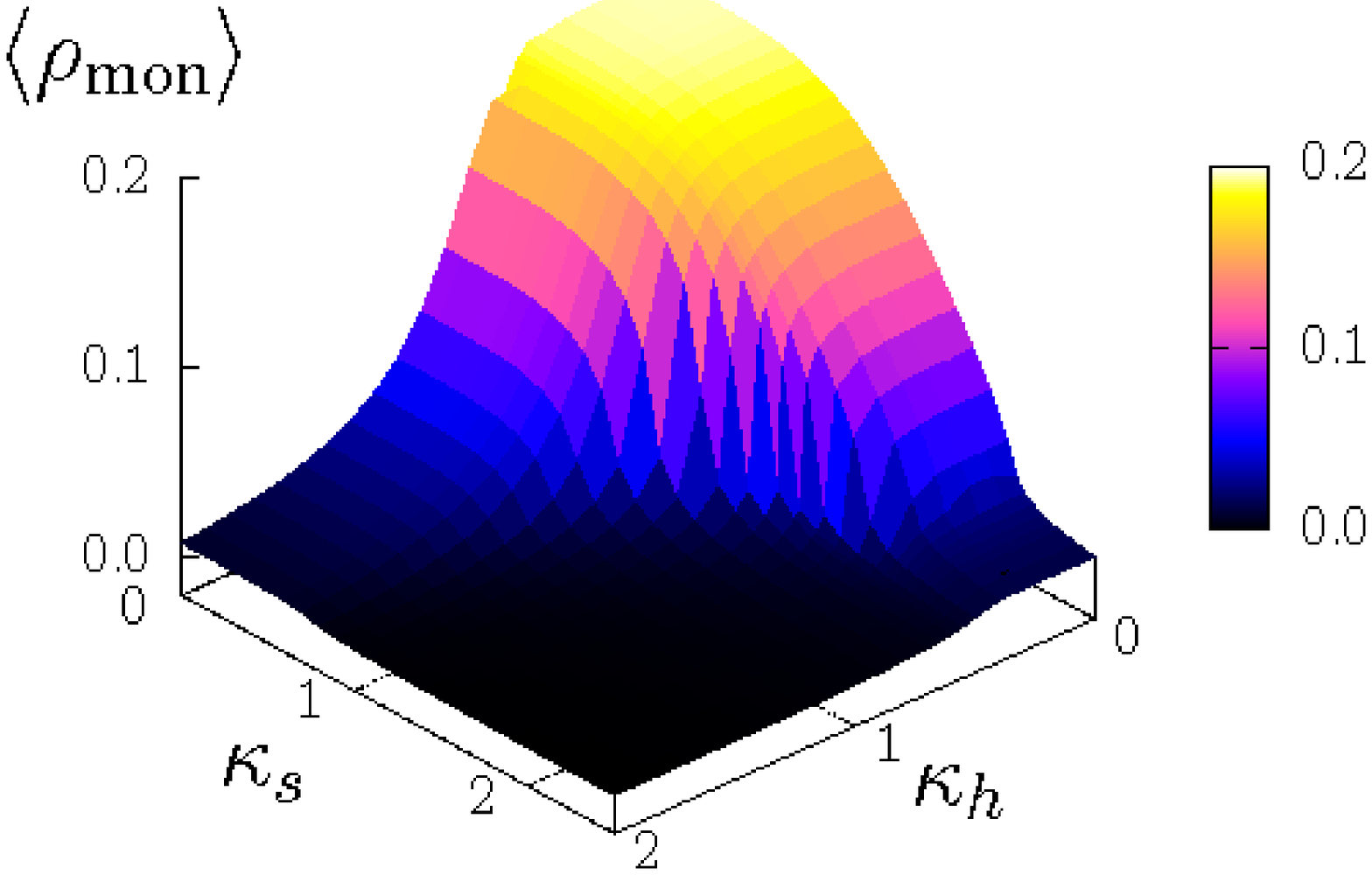}}\\
    \includegraphics[width=4.2cm]{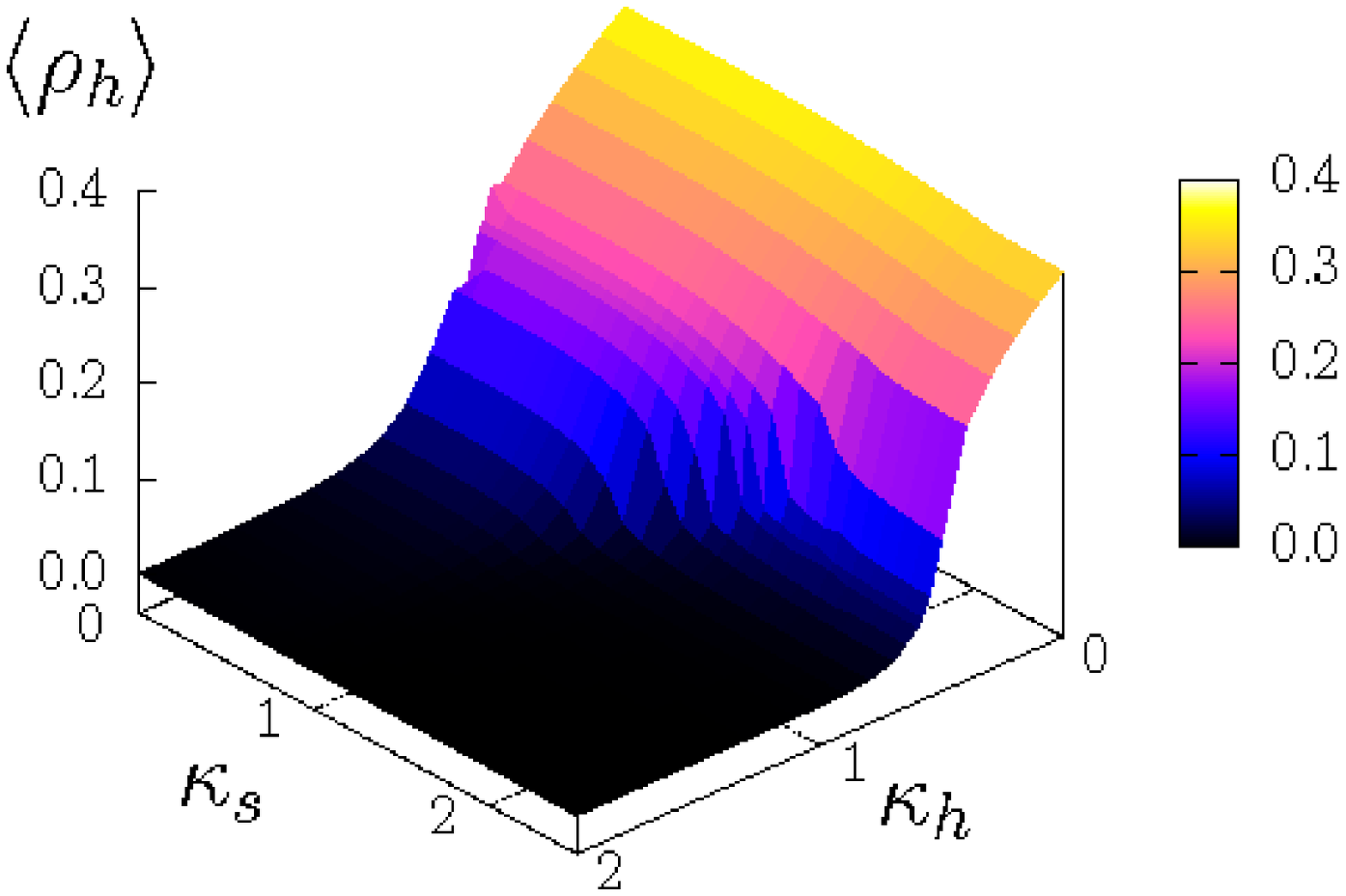} &
    \includegraphics[width=4.2cm]{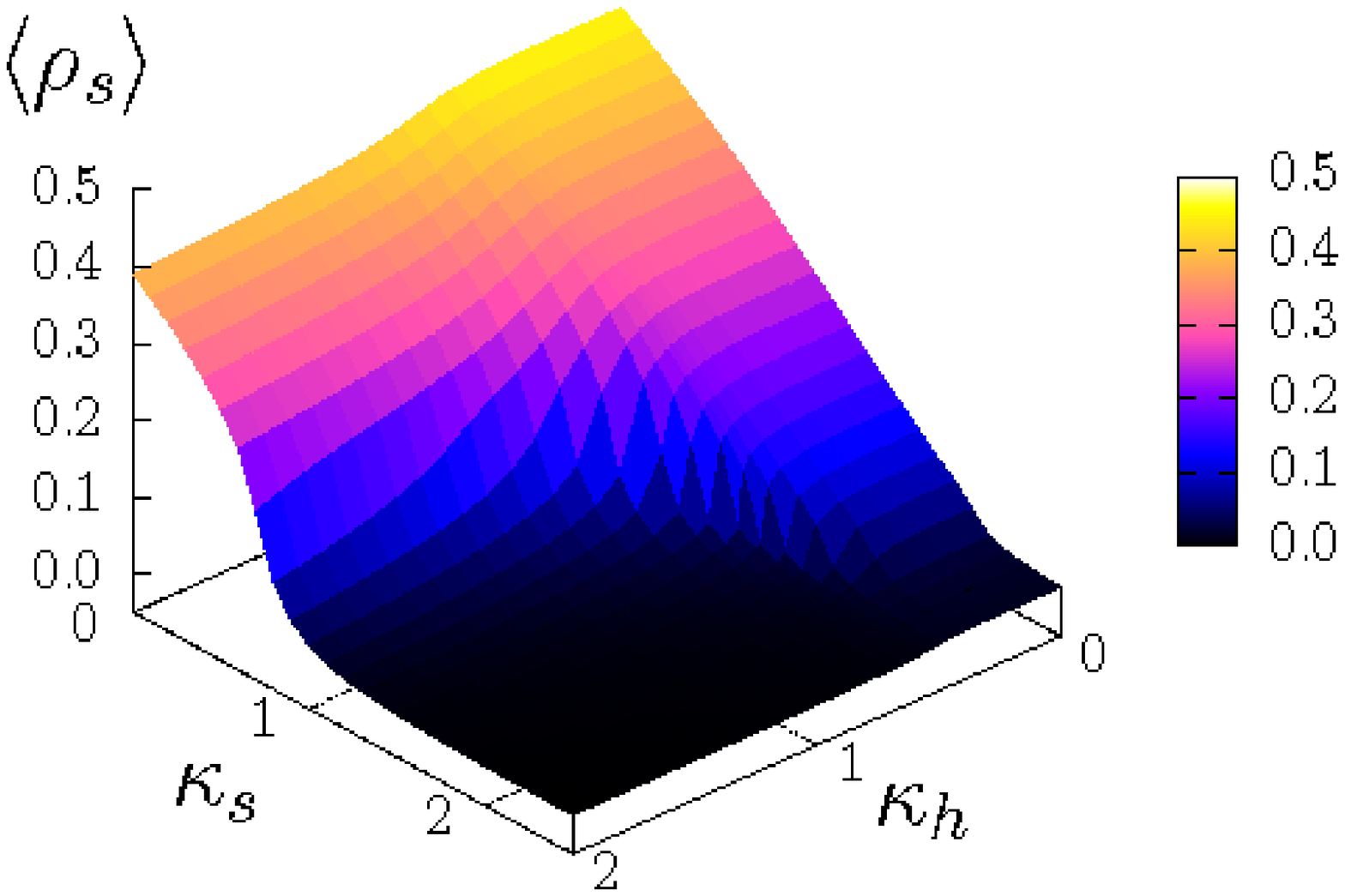}
  \end{tabular}
  \caption{The monopole density (top) and the vortex densities (bottom) at $\beta=1.0$ and $16^3$.
           The lower panel: the density of holon vortices (left) and of spinon vortices (right).}
  \label{fig:density:IR}
\end{figure}

In addition to the ``thermodynamical observables'' (\ref{eq:thdvariables}), the ``topological quantities'' (densities of defects) also show signals of a phase transition. The density of monopoles $\rho_{\rm mon}$ (at $\beta=1.0$ on a $16^3$ lattice) is plotted in the upper panel of Fig.~\ref{fig:density:IR}
over the $(\kappa_h$,$\kappa_s)$-plane. With increasing hopping parameters $\kappa_h$ or $\kappa_s$
the monopole density gets suppressed. As shown in the lower panels of Fig.~\ref{fig:density:IR}
the density of the holon vortices $\rho_h$ (spinon vortices $\rho_s$) significantly drops down with increasing $\kappa_h$ (or $\kappa_s$, respectively). This behavior is not unexpected because, as one can see from the vortex action~\eq{eq:vortex:int}-\eq{eq:vort:sh}, the larger the hopping parameter, the bigger the vortex mass. Therefore, the increase of a particular hopping parameter must suppress the density
of the corresponding vortex. As for the monopoles, the increase of either of the hopping parameters
should suppress the monopole density because the monopoles are connected by vortices according to
Eq.~\eq{eq:conservation} (see the example in Figure~\ref{fig:mon:conf}). The increase of tension (mass)
of either of the vortices leads to the confinement of the monopoles into magnetically neutral
monopole-anti-monopoles states, and, as a result, this leads to the suppression of the monopole density
as we see in Figure~\ref{fig:density:IR} (top). The maximum in the monopole density is seen
where both vortex densities are non-zero.

\subsection{Strengthening of the phase transition: ``2''+``2''=``1''}

The structure of the phase diagram at $\beta>\beta_c^{\mathrm{gI}}$ gives us the unique possibility to
study the phenomenon of merging (to be distinguished from crossing!) of two different phase transitions in a finite region of the coupling space.

In order to clarify the nature of the phase transitions, we studied the volume dependence of the average
of the plaquette and of the both link terms in the action~\eq{eq:cA2HM:action} as well as their respective susceptibilities in different regions of the phase diagram.
\begin{figure}[!htb]
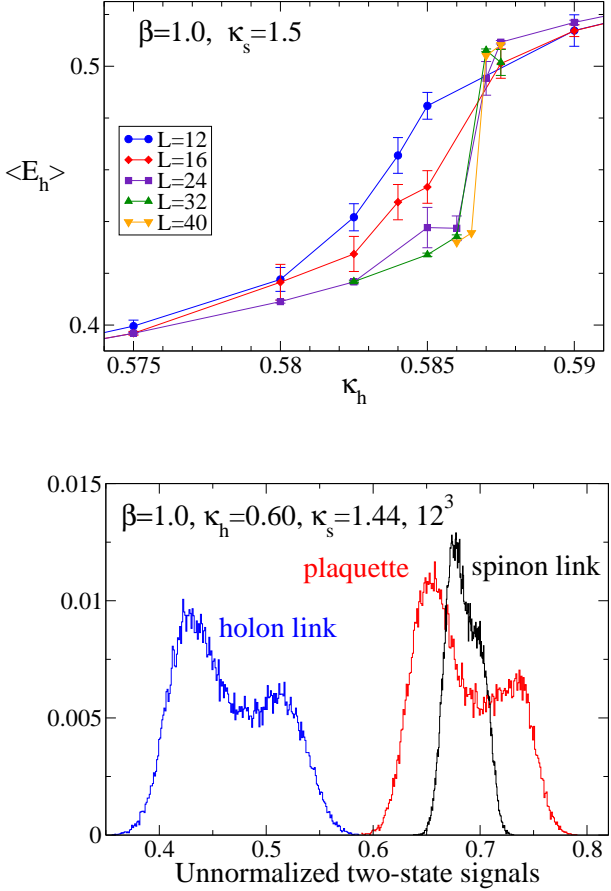

  \begin{tabular}{c}
  \hspace{-2mm}\includegraphics[width=8.0cm]{first_holonlink_k1_150_new.eps} \vspace{8mm}\\
  \hspace{2mm} \includegraphics[width=7.5cm]{hist_b10.eps}
  \end{tabular}
  \caption{Signatures for the first order transition at $\beta=1.0$: the holon link expectation value
           vs. $\kappa_h$ for different volumes at $\kappa_s=1.4$ (top), and (unnormalized) histograms
	   of the parts of the action in the pseudocritical region on a $12^3$ lattice (bottom).}
  \label{fig:thermodyn1}
\end{figure}
Figure~\ref{fig:thermodyn1}~(top) shows the jump developing in the holon link vs. $\kappa_h$ at fixed $\kappa_s$ with increasing volume. This jump is a clear signal of the first order nature of the transition
(it is actually observed in all parts of the action!) in the crossing region of the transition lines.
Note, however, that a lattice with a size $32^3$ turned out to be too large to tunnel for the selected $\kappa_h$ values, even within $ 5 \times 10^5$ Monte Carlo iterations. The reason is that the free energy barrier for such a large lattice is too high. In Figure~\ref{fig:thermodyn1} (bottom) we present typical two-state signals (here only for a $12^3$ lattice) of all three terms of the action at $(\kappa_h,\kappa_s)$ close to the transition.  The signals are strong along the {\it direct transition} line between the SM and SC phases. The two-state signals of the volume-averaged plaquette and volume-averaged holon link term become very weak when one goes to smaller $\kappa_h$ along the
horizontal dark-dotted (blue) line. Therefore in the crossing region between the two phases the strength of the phase transition is enhanced compared to the strength of the individual (horizontal and vertical)
transition lines. We discuss the order of these transition lines below.

It is known that for $\kappa_h \to 0$ the transition {\it vs.} $\kappa_s$ is of second order. Similarly, for large $\kappa_h$ the transition is most likely also of second order, again in the $XY$ universality class. We found that at $\kappa_h=2.0$ already for the largest volumes $40^3-48^3$ of our study the increase of the spinon link susceptibility stops as function of the lattice size.
This is the behavior expected for the $XY$ model at $\kappa_h \to \infty$~\cite{Campostrini:2000iw}.
Concentrating on two $\kappa_s$ values outside the crossing regime, where one of the transition lines (the light-dotted [red] one) runs vertical, we observe that there is no thermodynamic transition {\it vs.} $\kappa_h$ for the smaller $\kappa_s$. This is in agreement with what could be anticipated
for the limit $\kappa_s \to 0$.

Due to the weakness of the signals we were not able to check, whether for larger but finite values of $\beta$ (corresponding to weaker gauge coupling) there is still a line of first order transition somewhere
in the region of the phase diagram close to the crossing point. In fact, as $\beta$ increases the region in which the two transition lines join, tends to shrink.
Therefore the properties of the bulk observables -- probed by variations of the $\kappa_h$
hopping parameter -- provide a weak signal if the sequence of measurements does not pass through
the right (crossing) point at the right (corresponding to a maximal variation of the bulk
observables) angle in the coupling space. Anyway, the first order phase transition should inevitably
disappear from the crossing region in the region $\beta\to\infty$ as we surely know from
Figure~\ref{fig:phase:2D:VI}.

\begin{figure}[!htb]
  \begin{center}
  \includegraphics[width=7.5cm]{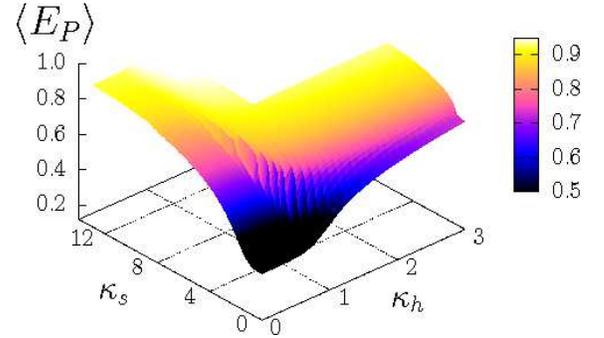}
  \end{center}
  \caption{Plaquette expectation value in the ($\kappa_h,\kappa_s$)-plane at $\beta=0.8$ on a $16^3$
           lattice.}
  \label{fig:act_plaq_08}
\end{figure}

Increasing further the strength of the gauge coupling (decreasing $\beta$ below $\beta=1.0$) one should finally approach the critical value of the gauge Ising model, $\beta_c^{\rm{gI}}\approx 0.7613$, below which the structure of the two-dimensional phase diagram qualitatively changes. In order to see what happens there we studied in detail the cross-section of the phase diagram at $\beta=0.8$ which is already
very close to the critical value. In Figure~\ref{fig:act_plaq_08}
we present the plaquette expectation value $\langle E_P\rangle$ in the ($\kappa_h$,$\kappa_s$)-plane. The white areas in the Figure correspond to uninteresting regions in the plane of the couplings at simultaneously large $\kappa_h$ and $\kappa_s$ values. From this Figure we may expect that the phase transition line corresponding to the spinon condensation is parallel to the $\kappa_h$ axis for not too small $\kappa_s$. The same effect can also be seen in the landscape of the average spinon action $\langle E_s\rangle$ (not shown). Analogously, the holon condensation line is parallel to the $\kappa_s$ axis for not too small $\kappa_h$ as can also be seen in the landscape of the average holon action
$\langle E_h\rangle$ (not shown).

\begin{figure}[!htb]
  \begin{tabular}{c}
  \includegraphics[width=7.5cm]{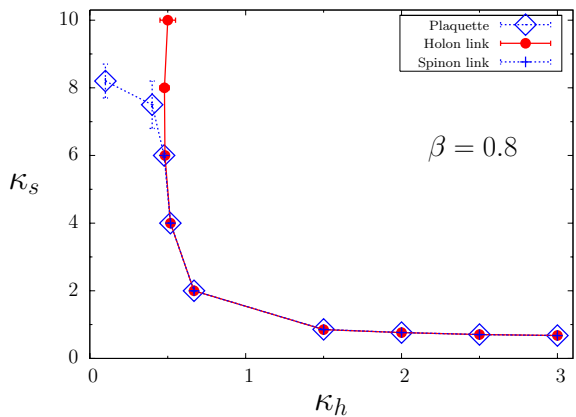}
  \end{tabular}
  \caption{The phase transition lines in the ($\kappa_h$,$\kappa_s$)-plane
           obtained by searching for peaks of various susceptibilities
           using multihistogram reweighting for $\beta=0.8$ on the $16^3$ lattice.}
  \label{fig:thermodyn08}
\end{figure}

The maxima of the susceptibilities of the different action contributions demonstrate in more detail
the exact location of the phase transition curves. To obtain accurate pseudocritical couplings we performed several high statistics runs near chosen points indicated in Figures~\ref{fig:thermodyn08} changing one of the hopping parameters to cross the expected phase transition line.

All obtained Monte-Carlo histories of measurements have been evaluated together using a multihistogram reweighting procedure~\cite{ref:FS}. The combined histograms of observables at a chosen pair of couplings $\kappa_h,\kappa_s$ (interpolating among the $\kappa_h,\kappa_s$-grid of the used data sets) are obtained by reweighting from the closest data points in the grid. They contain all information necessary to precisely locate the phase transition. In particular, using such a combined histogram interpolating in
the range of the chosen hopping parameters we were able to identify the pseudocritical value where the reweighted histogram exhibits a maximum in the susceptibility of the corresponding variable.

To estimate the error in determining the critical couplings, we blocked our original data from several couplings into blocks similar to a jackknife method and constructed individual ``subhistograms'' at the chosen coupling pair for those data subsets. Using those subhistograms to find the maximum of the susceptibility, different critical couplings have been found which allowed to estimate the accuracy of the location of the critical coupling using all available data.

The results of this detailed investigation is shown in Figures~\ref{fig:thermodyn08}.
We used thermodynamical variables in the reweighting as indicated in the legend. Most errors in the pseudocritical couplings are smaller than the symbol sizes, the lines are drawn to guide the eyes.

The individual observables show common signals in certain parts of the phase plane. The region of
the common transition line has been increased compared to results at larger $\beta$. In addition to the gross structure the susceptibility of the plaquette action shows a weak but measurable signal for the continuation of the ``horizontal'' line at large $\kappa_s^{\rm {crit}}\approx 8.0$ for very small $\kappa_h$.

\begin{figure}[!htb]
  \begin{tabular}{c}
    \includegraphics[width=7.5cm]{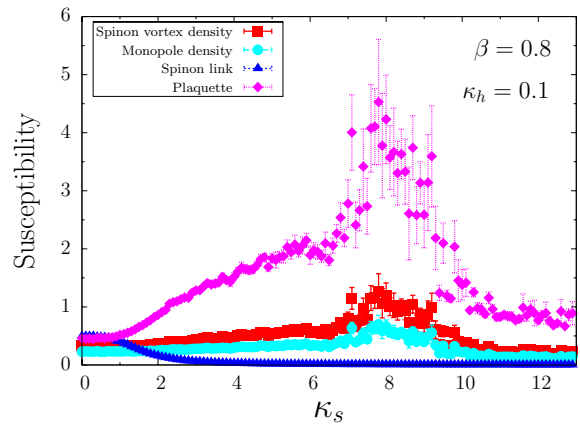}
  \end{tabular}
  \caption{The susceptibilities of thermodynamical and topological quantities
           at $\beta=0.8$, and $16^3$ as function of $\kappa_s$ at $\kappa_h=0.1$.}
  \label{fig:susc_08_k1eq01}
\end{figure}

This has been studied in more detail at small $\kappa_h=0.1$ in Figure~\ref{fig:susc_08_k1eq01}.
Weak but relatively sharp signals are observed in some of the susceptibilities shown as functions of the spinon hopping parameter $\kappa_s$. Note that contrary to the spinon link susceptibility, a signal is present in the susceptibility of the spinon vortex density. The observed behavior can be interpreted as indication for the expected second order phase transition.

The emerging picture is in agreement with our expectation, that the critical value of the complementary $\kappa_{s}$ should take an infinite value as the value of $\beta$ approaches the critical coupling $\beta_c^{\rm gI}$ of the gauge Ising model. Indeed, there is no transition at finite $\kappa_s$ for the $Q=2$ Abelian Higgs model at $\beta_c^{\rm gI}$.

Figure~\ref{fig:thermodyn08} shows the location of the phase transition lines in a large region of the ($\kappa_h$,$\kappa_s$)-plane. In particular we are convinced that the holon and the spinon-pair condensation lines join for strong gauge couplings just above the gauge Ising model's critical value.

\begin{figure}[!htb]
  \begin{center}
  \includegraphics[width=7.5cm]{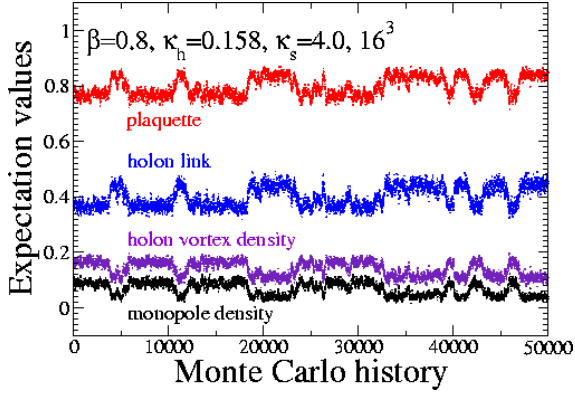}
  \end{center}
  \caption{Two-state signals in the Monte Carlo history of several observables near the center
           ($\kappa_s=4.0$) of the merging line of the two transition line as observed at $\beta=0.8$
	   on a $16^3$ lattice.}
  \label{fig:history}
\end{figure}

In order to demonstrate that at $\beta=0.8$ the phase transition in the merging region is of first order
we have considered three different fixed $\kappa_s$ values, $\kappa_s=2.0$, 4.0 and 6.0.
In Figure~\ref{fig:history} we present for $\kappa_s=4.0$ near criticality nice two-state signals visible in various observables, indicating that the systems jumps from one metastable state to another. This is a clear characteristic of a first order phase transition.
The signal is strongest near the center value $\kappa_s=4.0$, and then becomes weaker for smaller and larger $\kappa_s$ values (not shown) where -- according to our expectations -- the nature of the transition gets closer to a second order transition.

\begin{figure}[!htb]
  \includegraphics[width=7.5cm]{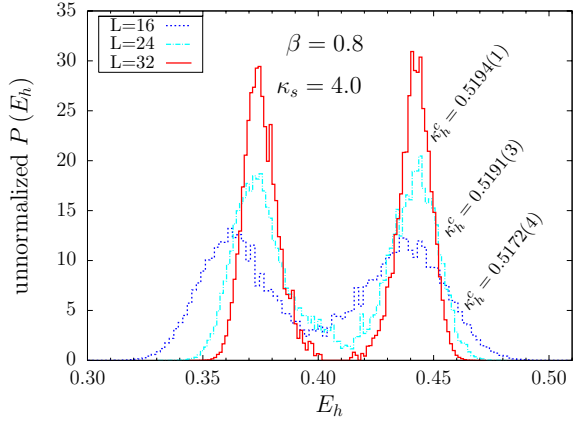}
  \caption{The two-state signal in the histogram of the holon link expectation value $E_h= S_h/N_L$
           at $\beta=0.8$ and $\kappa_s=4.0$ with varying lattice volume reweighted to the transition
	   value $\kappa_h^c$ at maximal susceptibility.}
  \label{fig:twostateE_h:0.8}
\end{figure}

The tunnelings between different vacua in the region of the first order phase transition are clearly visible for small lattice volumes (i.e., $16^3$). They can also be seen in the (unnormalized)
distribution of the holon energy term shown in Figure~\ref{fig:twostateE_h:0.8} for different lattice volumes and various values of the holon hopping parameter.

\begin{figure}[!htb]
  \begin{tabular}{c}
    \includegraphics[width=7.5cm]{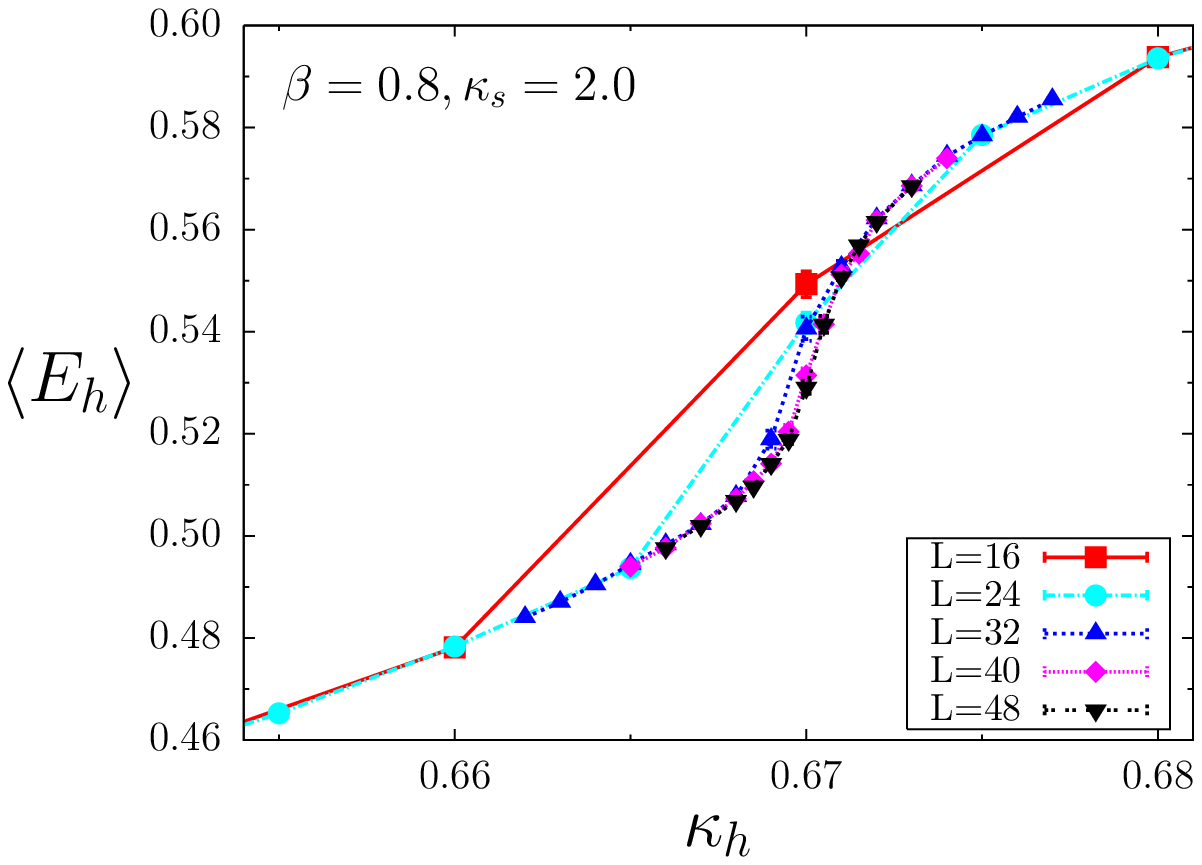} \\
    \includegraphics[width=7.5cm]{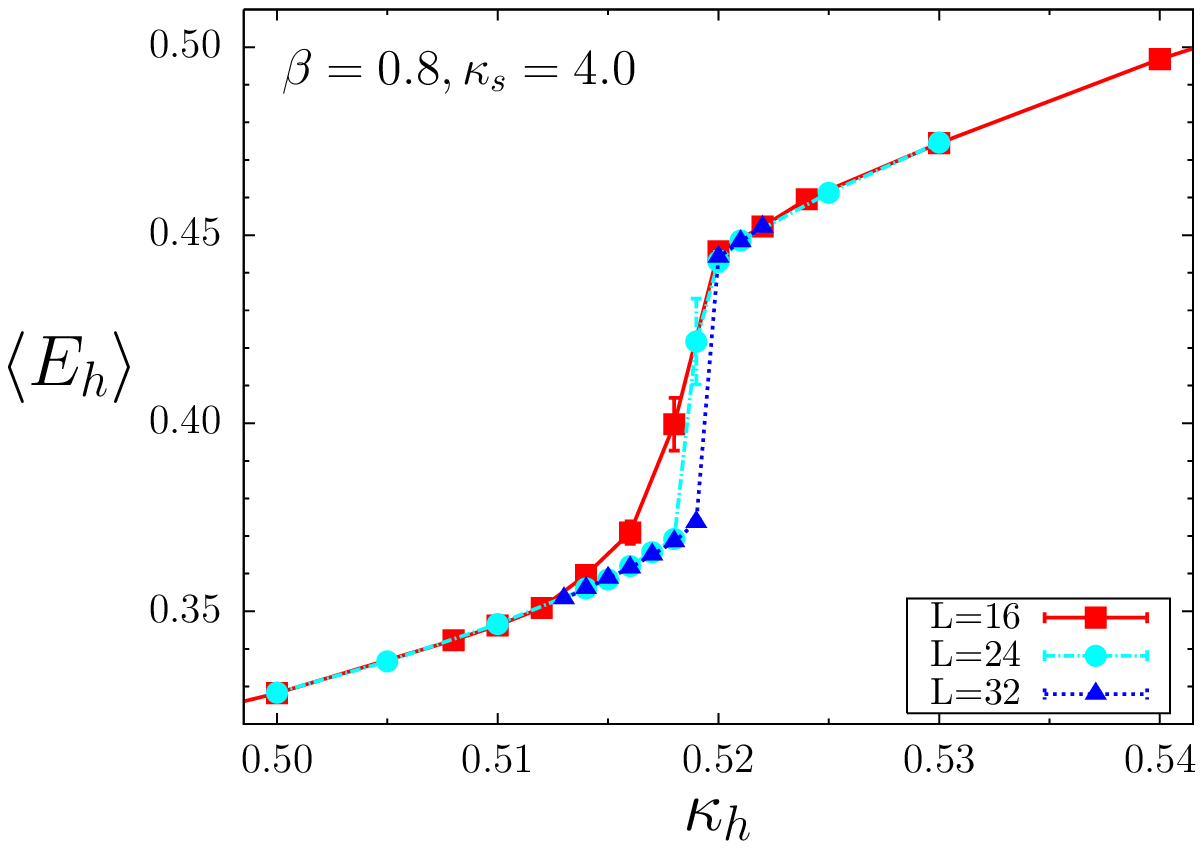} \\
    \includegraphics[width=7.5cm]{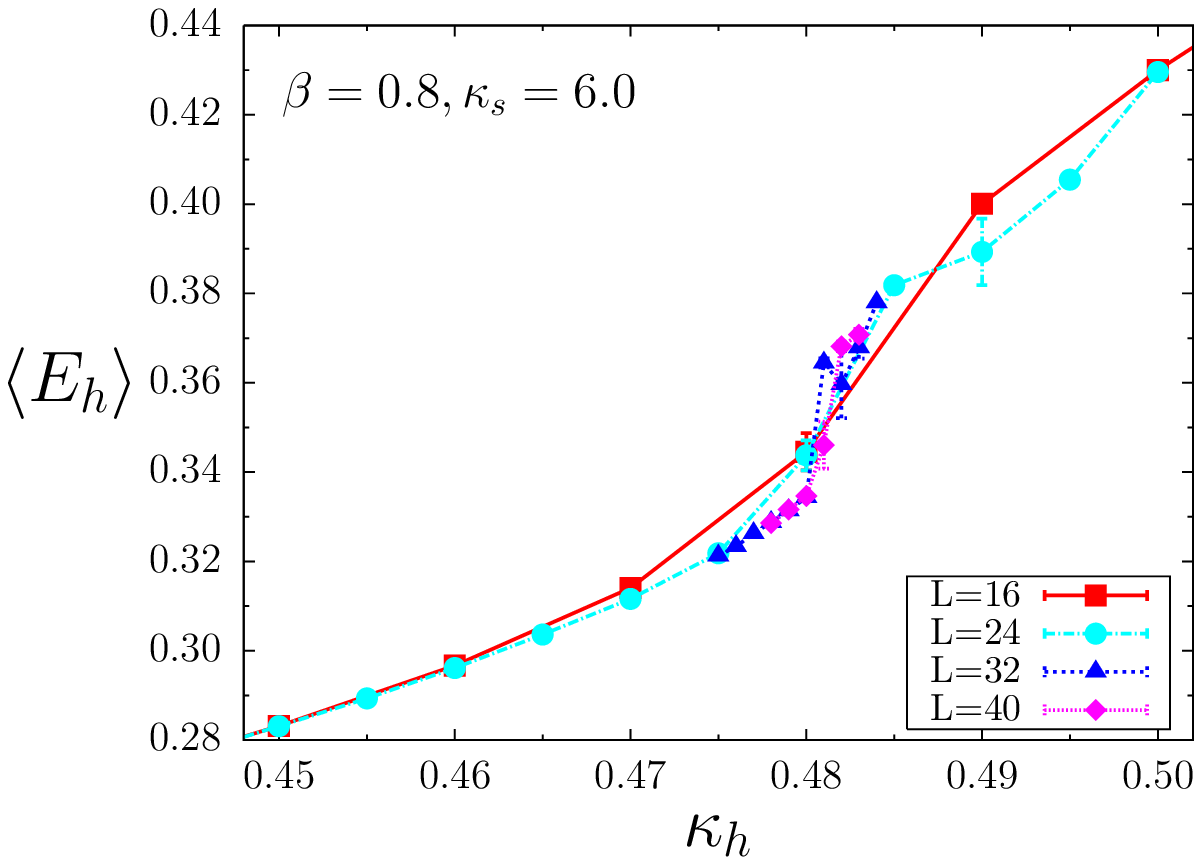}
   \end{tabular}
  \caption{Holon link energy expectation value as a function $\kappa_h$ for various volumes at
           $\beta=0.8$ for $\kappa_s=2.0$ (top), $\kappa_s=4.0$ (middle) and $\kappa_s=6.0$ (bottom).}
  \label{fig:plaq_08_k2fixed}
\end{figure}

The typical features of the first order phase transition can also be observed, for example, in the behavior of the holon link expectation value. This observable -- shown in Figures~\ref{fig:plaq_08_k2fixed} as a function of $\kappa_h$ for the same three chosen values of the hopping parameter $\kappa_s$ -- clearly develops a jump at the phase transition point as the volume of the lattice increases. Thus the model has a non-vanishing latent heat at the phase transition point characterizing a first order phase transition.
The signal of the first order is most clearly visible at $\kappa_s=4.0$, while it becomes weaker at larger and smaller values of $\kappa_s$ in agreement with our observations made with the help of the Monte Carlo
histories.

\begin{figure}[!htb]
  \begin{tabular}{c}
    \includegraphics[width=7.5cm]{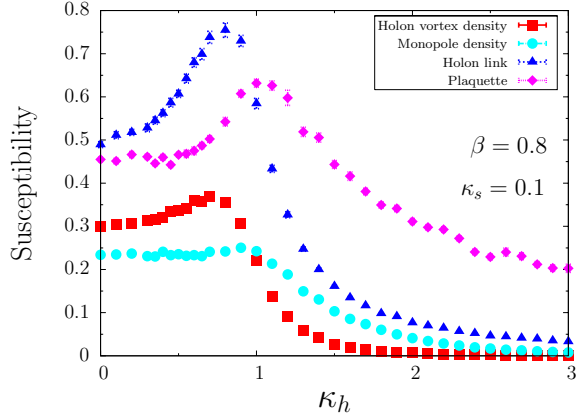} \\
  \end{tabular}
  \caption{The susceptibilities of thermodynamical and topological quantities
           at $\beta=0.8$, and $16^3$ as function of $\kappa_h$ at $\kappa_s=0.1$.}
  \label{fig:susc_08_k2eq01}
\end{figure}

In order to confirm the absence of the phase transition as we expect from either Figure~\ref{fig:phase:2D:section1}~(a) or Figure~\ref{fig:phase:2D:section1}~(c), we investigated the
model numerically for a much smaller value of the spinon hopping parameter, $\kappa_s=0.1$. The crossover nature of the transition is clear from the susceptibilities of the thermodynamical and topological
observables (in Figure~\ref{fig:susc_08_k2eq01}): we observe either no susceptibility signal or broad maxima at different values of $\kappa_h$.
Of course, a remnant of a vortex percolation transition -- expected at small $\kappa_s$ -- cannot be ruled out by our investigations and has to be studied using cluster analysis techniques.

Thus, we have strong reasons to conclude that the phase transition scenario at relatively strong gauge coupling, but still above the critical $\beta^{\mathrm{gI}}_c$ of the gauge Ising model, resembles more
one of the anticipated scenarios, the phase structure of Figure~\ref{fig:phase:2D:section1}~(a)
rather than the closest alternative scenario plotted in Figure~\ref{fig:phase:2D:section1}~(c).

Summarizing, we have clearly observed the strengthening of the phase transition in the merging region.
The strengthening happens due to activity of the internal gauge field in the weak coupling region.

\subsection{The phase structure at strong coupling below~$\beta^{\mathrm{gI}}_c$}

In this Section we identify the phase structure of the model choosing a really strong gauge coupling, $\beta=0.5 < \beta^{\mathrm{gI}}_c$. As in the previous subsections we first make an exploratory study of the phase structure for a relatively small lattice size, $16^3$.

\begin{figure}[!htb]
  \begin{tabular}{cc}
    \includegraphics[width=4.2cm]{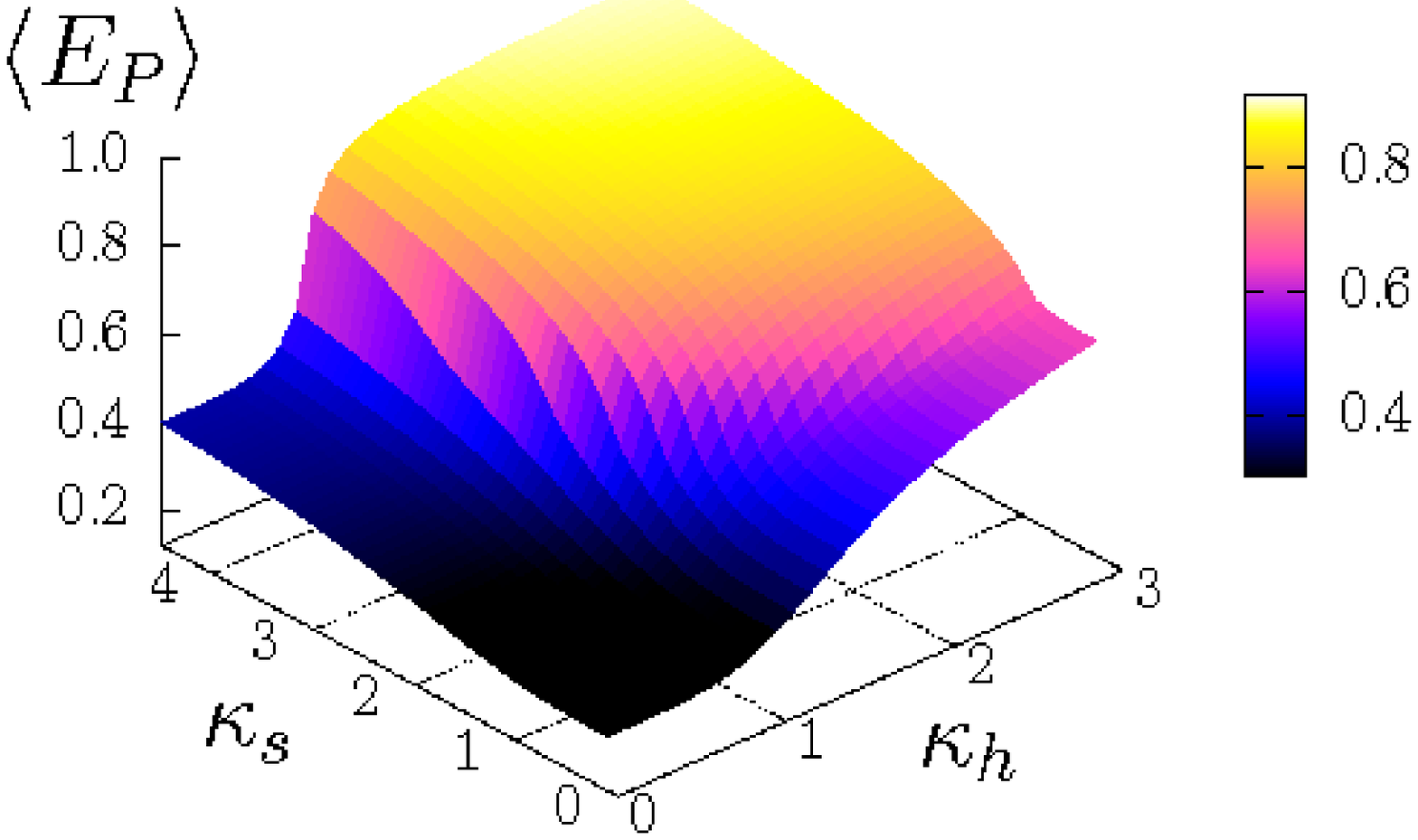} &
    \includegraphics[width=4.2cm]{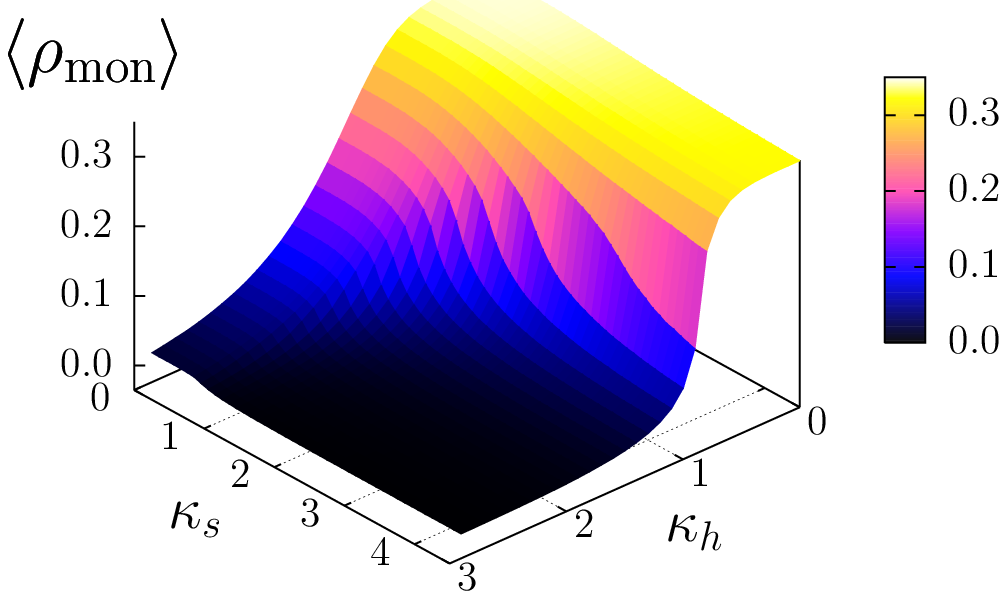} \\
    \includegraphics[width=4.2cm]{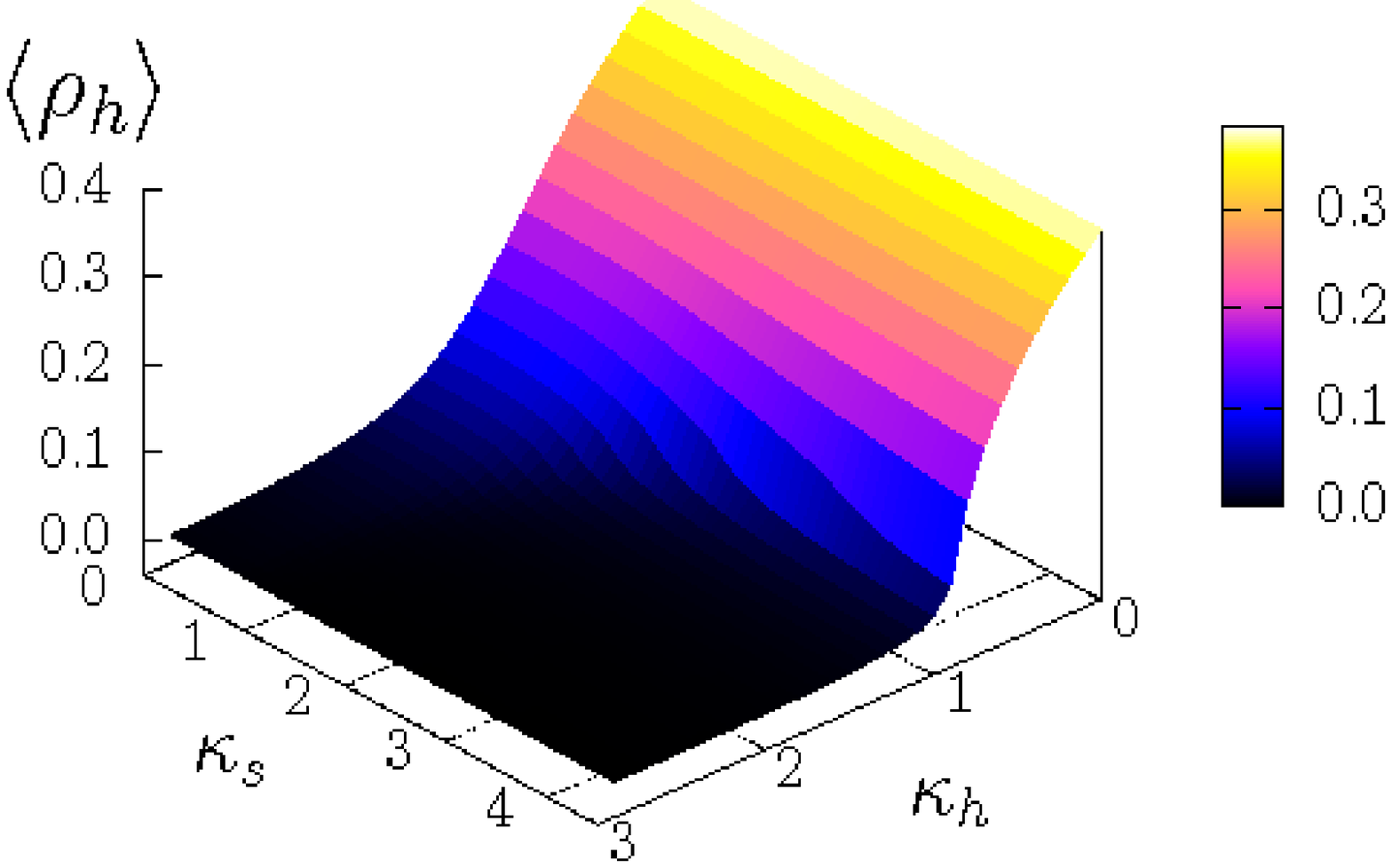} &
    \includegraphics[width=4.2cm]{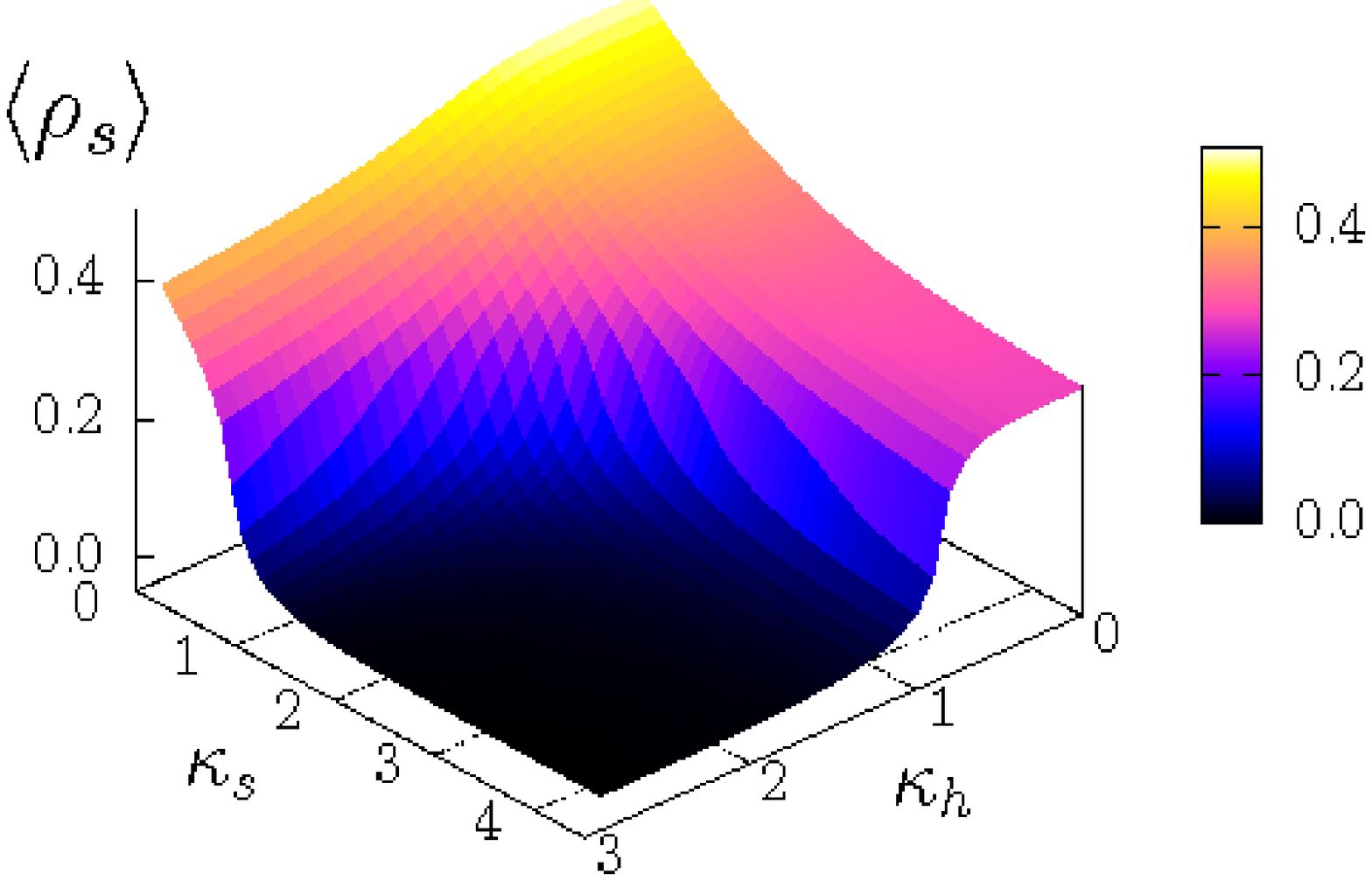}
   \end{tabular}
  \caption{Expectation value of the gauge plaquette and the densities of the topological defects
           (monopoles, holon vortices and spinon vortices) at $\beta=0.5$ on a $16^3$ lattice.}
  \label{fig:dens_05}
\end{figure}

In Figures~\ref{fig:dens_05} we show the plaquette expectation value and the average densities of all topological defects in the $(\kappa_h,\kappa_s)$-plane (for the topological densities the viewpoint is rotated for convenience).
Using the gauge degrees of freedom, the approximate phase structure is most easily detected.

It is interesting to discuss the density of the topological defects since they may provide us with the hint how the condensates behave in the different regions of the phase diagram (a condensate is suppressed if the density of the corresponding vortex is non-zero). The density of the holon vortices behaves qualitatively similarly to the one in the weak coupling case, Figure~\ref{fig:density:IR}. In contrast to this, the density of the spinon vortices is qualitatively different at strong and weak gauge couplings as a function of $\kappa_s$ and $\kappa_h$: while in the weak coupling regime the spinon vortex density rapidly
vanishes going to the large-$\kappa_s$ limit, the same is no longer true in the strong coupling regime.
The monopole density is also different in both regimes: in the weak coupling regime the monopole density
is visibly non-zero only in the vicinity of the $\kappa_h,\kappa_s$ origin, while in the strong coupling regime the monopole density remains almost unaffected towards large values of $\kappa_s$. Note that similarly to Figure~\ref{fig:density:IR} the rule remains true that a non-zero monopole density
is observed only in the region of the phase diagram where both vortex densities are non-vanishing.

Figures~\ref{fig:dens_05} show approximately a structure of the phase diagram qualitatively consistent with our expectations plotted in Figure~\ref{fig:phase:2D:section1}~(b).

\begin{figure}[!htb]
  \begin{tabular}{c}
  \includegraphics[width=7.5cm]{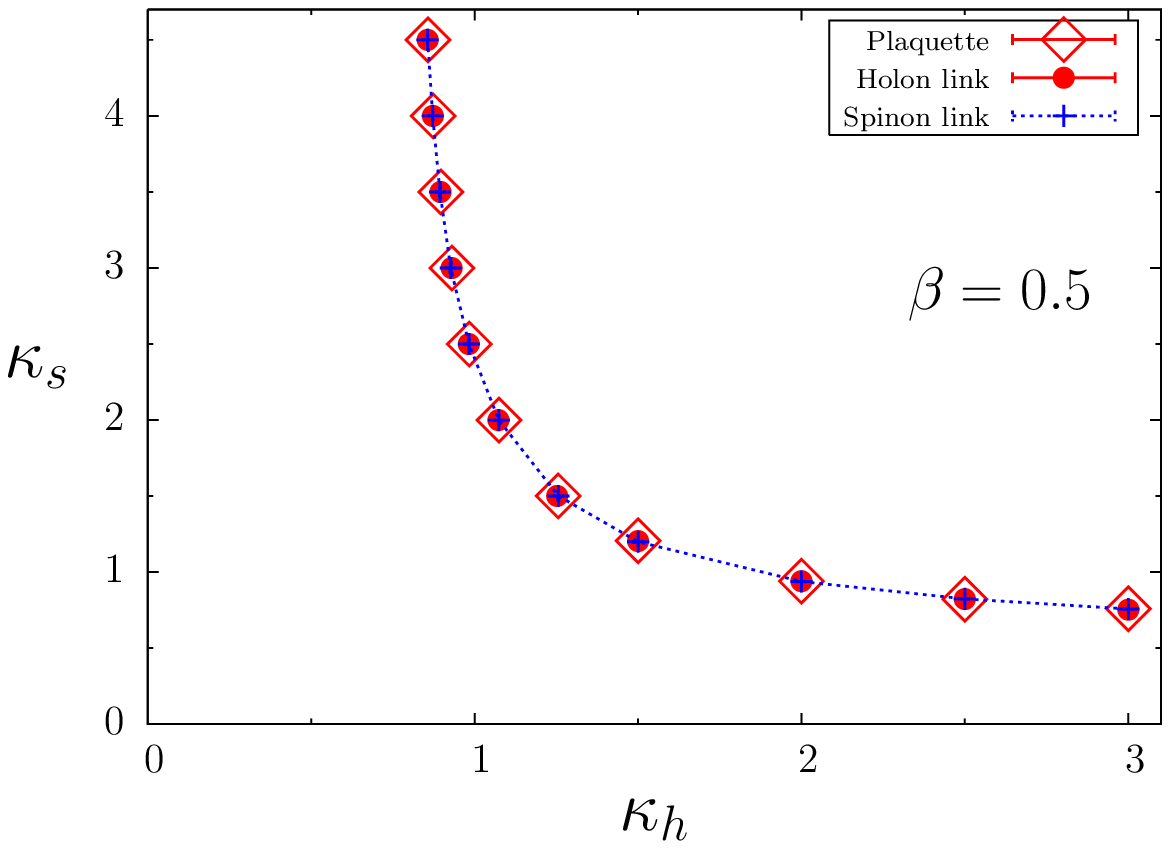} \\
  \includegraphics[width=7.5cm]{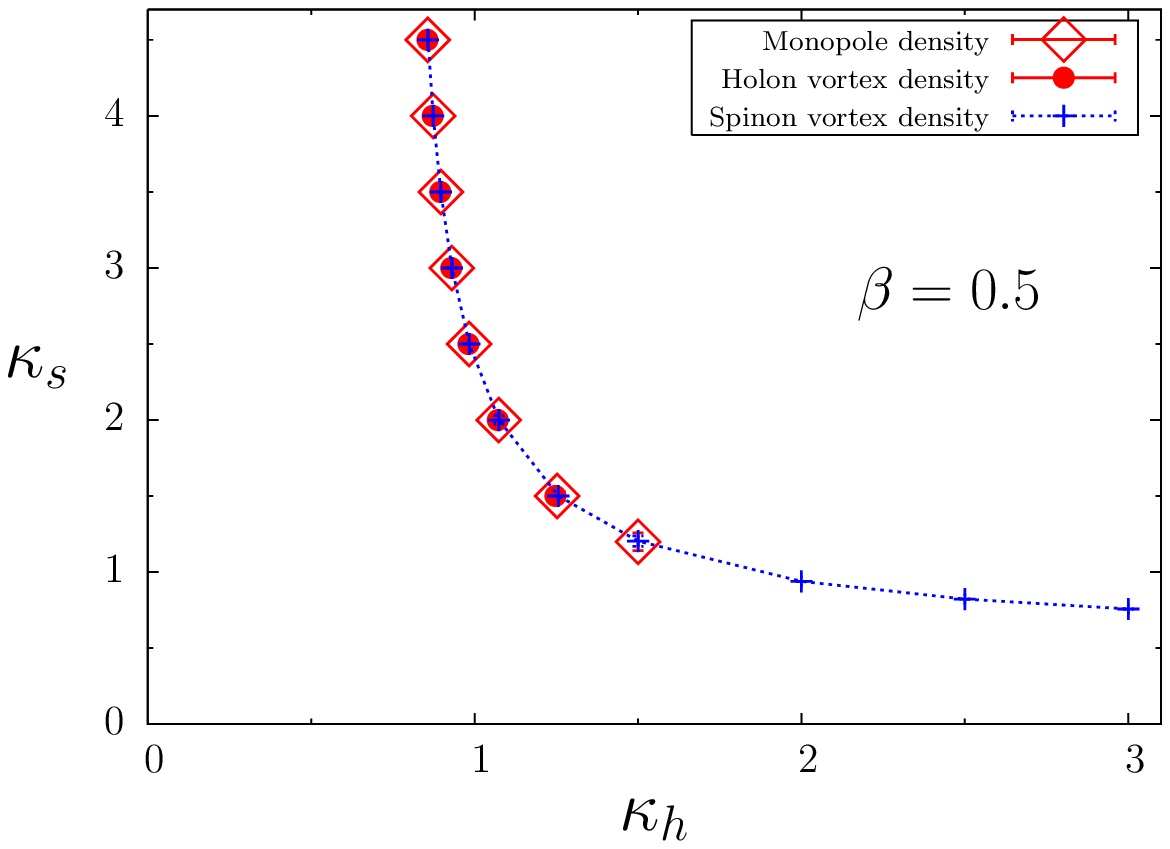}
  \end{tabular}
  \caption{The phase transition line in the ($\kappa_h$,$\kappa_s$)-plane at fixed $\beta=0.5$
           on the lattice $16^3$ to be located by finding the susceptibility peaks of
	   thermodynamical (top) and topological (bottom) variables.}
  \label{fig:phase:0.5}
\end{figure}

In order to study the location of the phase transition line more precisely, we have studied the thermodynamical and topological susceptibilities interpolated by reweighting techniques as for
$\beta=0.8$. The result of a search for maxima of those susceptibilities is represented in Figures~\ref{fig:phase:0.5}.

We note that the different variables mark a single line in the hopping parameter plane (within our resolution). Signals from the thermodynamical susceptibilities (upper Figure~\ref{fig:phase:0.5}) are
found in the whole considered ($\kappa_h,\kappa_s$)-range. In contrary, using the topological observables (lower Figure~\ref{fig:phase:0.5}), only the maximum of the spinon vertex susceptibility $\chi_{{}_{\rho_s}}$ signals the transition along the whole line. The monopole and holon vertex signals
are seen in the common ``vertical'' line only.

\begin{figure}[!htb]
  \begin{tabular}{c}
  \includegraphics[width=7.5cm]{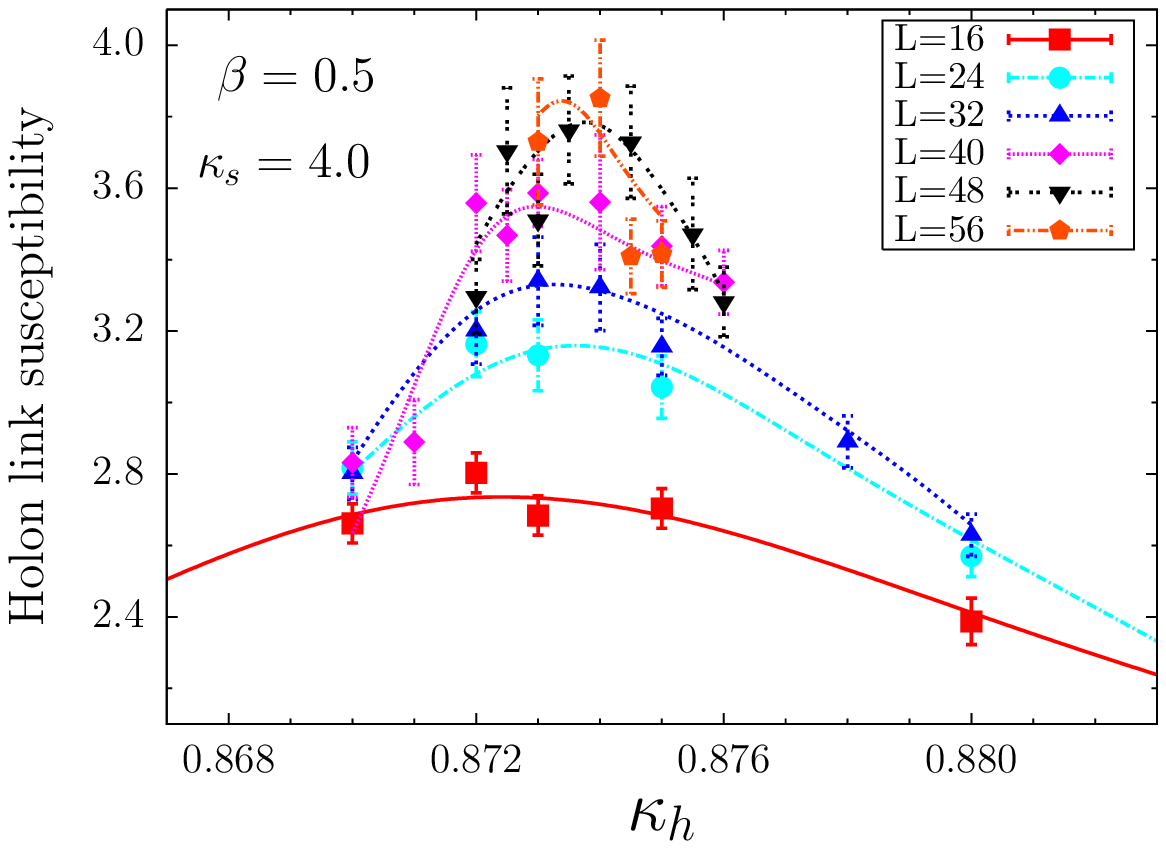}\\
  \includegraphics[width=7.5cm]{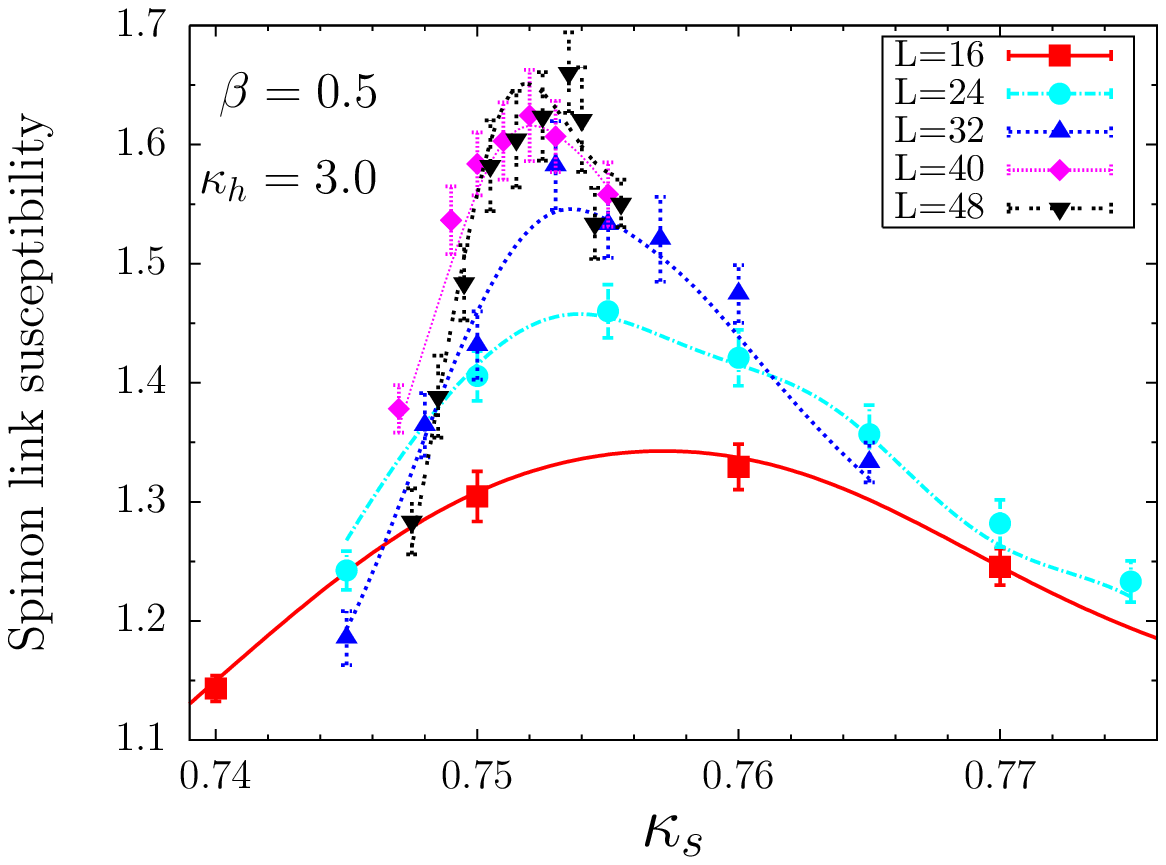}
  \end{tabular}
  \caption{The holon link susceptibility vs. $\kappa_h$ at fixed $\kappa_s=4.0$ (top) and the spinon
           link susceptibility vs. $\kappa_s$ at fixed $\kappa_h=3.0$ (bottom) for various lattice
	   volumes at $\beta=0.5$ together with the curves obtained by multihistogram reweighting.}
  \label{fig:susc_link_05_fixed}
\end{figure}

The volume dependence of the susceptibilities of the holon link (top) and the spinon link (bottom) along lines crossing the phase transition line horizontally at $\kappa_s=4.0$ and vertically at $\kappa_h=3.0$
definitely rules out the possibility of a first order transition (Figure~\ref{fig:susc_link_05_fixed}).

\begin{figure}[!htb]
  \includegraphics[width=7.5cm]{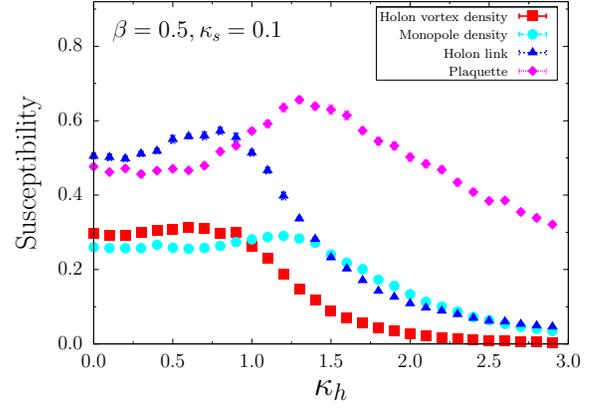}
  \caption{Susceptibilities of thermodynamical and topological quantities vs. $\kappa_h$
           at $\beta=0.5$, $\kappa_s=0.1$ and $16^3$.}
  \label{fig:susc_05_k2eq01}
\end{figure}

Similarly to the weak coupling case $\beta> \beta_c^{\rm {gI}}$, a branch of phase transition is absent ranging to very small $\kappa_s=0.1$, as it can also be seen from the behavior of various susceptibilities,
in Figure~\ref{fig:susc_05_k2eq01}. Indeed, these susceptibilities develop non-sharp maxima at
significantly different~$\kappa_h$. Note that the existence of a remnant from a vortex percolation
transition cannot be determined by our calculations. It has to be sought for by cluster analyses.

Summarizing this subsection we mention in short, that we observe one main phase transition line which is most presumably of second order. In addition, we see a very weak signal along the vertical line $\kappa_h \sim 1$ for small $\kappa_s \leq 1.0$ intersecting the $\kappa_s=0$ axis. However, this mentioned signal is not a thermodynamical transition.

The observed picture at $\beta=0.5$ is consistent with the phase diagram proposed in Figure~\ref{fig:phase:2D:section1}~(b). We clearly observe the common line H-B' indicating the $\Delta$-condensation as signaled by the susceptibility maximum of the spinon vortex density and the behavior of the density itself. The ``vertical'' section H-F' of the line H-B' -- showing in addition the $b$-condensation -- is also present as indicated by the behavior of the holon vortex density.

There is no hint for thermodynamically separated transitions for the condensations. If a line E-F' with a point E exists, it could be related to percolation properties as can be conjectured from Figures~\ref{fig:dens_05} and \ref{fig:susc_05_k2eq01}.

\section{The strengthening of phase transitions and a possible realization in QCD}
\label{sec:QCD}

This paper is devoted to the investigation of non-trivial properties of a model containing two Higgs fields coupled to a single compact gauge field. As most remarkable observation we find the possibility that
two weak (second order) phase transitions -- marking the vanishing/non-vanishing of two different condensates -- tend to join and constitute a stronger (first order) transition at sufficiently strong (but not too strong) gauge coupling. Although the model is related to strongly correlated electron systems
in (2+1) dimensions (at zero temperature), we conjecture that the observed effect is much more general such that it can also be realized in the field theory of strong interactions, Quantum Chromodynamics (QCD).
The two-dimensional phase diagram that attracts the attention of lattice theorists and phenomenologists
today, is spanned by temperature and non-zero baryonic density.

\begin{figure}[!htb]
  \begin{center}
  \includegraphics[width=7.5cm]{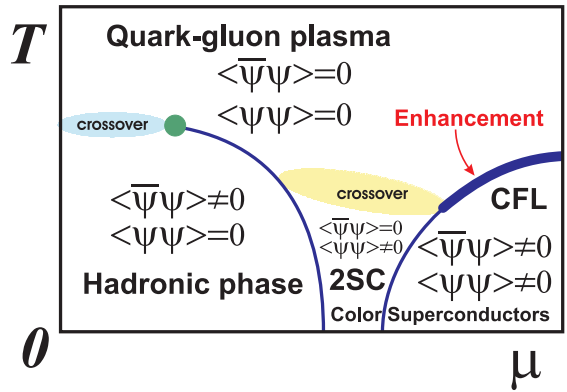}
  \end{center}
  \caption{Schematic view of the phase diagram of QCD with a suggested enhancement to a first order
           phase transition due to the mediating, strongly coupled gauge field. The explanation is
	   given in the text.}
  \label{fig:QCD}
\end{figure}

The schematic view of the QCD phase diagram in terms of the baryon chemical potential $\mu$ and the temperature $T$ is shown in Figure~\ref{fig:QCD}.
For a detailed review we refer to Refs.~\onlinecite{ref:Alford,ref:Stephanov}. The diagram corresponds to
QCD with two light quarks ($u$ and $d$) and one heavier quark ($s$). The order parameters which are relevant in this phase diagram are the chiral condensate, $\langle \bar\psi \psi\rangle$ and the diquark condensate, $\langle \psi\psi\rangle$. It is worth mentioning that due to the presence of three quark flavors the phase structure of the realistic QCD is richer compared to Figure~\ref{fig:QCD}. The structure of the superconducting phases is finer because there are three (instead of one) diquark condensates related to different flavors, and the phases with different combination of non-zero diquark condensates may emerge. A thorough treatment of the QCD phase diagram in an effective model of the quark matter, known as
the Nambu-Jona-Lasinio (NJL) model~\cite{ref:NJL}, can be found in Ref.~\onlinecite{Ruster:2005jc}.
In view of the qualitative character of our considerations we ignore the color, flavor and Dirac indices in the notation, and we do not discriminate between different quark flavors taking the schematic picture
shown in Figure~\ref{fig:QCD} as a reference point.

At low temperature and density, strongly interacting matter is in the hadronic phase which is characterized by broken chiral and unbroken color symmetries, so that the quark condensate is non-vanishing while the diquark condensate vanishes. As temperature increases, matter goes over into the quark-gluon plasma (QGP) phase in which both condensates vanish. At small chemical potential the transition between these phases is
probably a smooth crossover~\cite{ref:qcd:crossover} rather than a real phase transition, so that in this region the difference between the two phases is marginal (and the vanishing of
$\langle \bar\psi \psi\rangle$ is not complete). However, the chiral condensate is drastically suppressed
in the high temperature regime, what we formally express in Figure~\ref{fig:QCD} by
$\langle \bar\psi \psi\rangle=0$. We adopt the same convention for other exponentially suppressed but non-vanishing condensates. The phases on both sides of a crossover are often said to be connected via an
analytical continuation. Consequently, in a strict mathematical sense, the chiral condensate does not take zero value at any point of the quark-gluon plasma phase.

As the chemical potential increases, the crossover between the hadron and the QGP phases turns into a second order transition denoted by a dot in Figure~\ref{fig:QCD}. The second order phase transition is an endpoint of a first order phase transition line that separates the hadronic phase not only from the QGP phase but also from one of the color superconducting phases, the so called 2SC phase, in which two colors are locked with the global flavor symmetries through the formation of a non-zero diquark condensate,
$\langle \psi \psi\rangle \neq 0$. Note, that in the 2SC phase the chiral condensate is zero~\cite{ref:Alford}.

The 2SC phase is presumably also separated by a crossover from the QGP phase and by a first order phase transition from another color superconducting phase in which all three colors are locked with all three flavor degrees of freedom (this phase is known as ``color-flavor locked'', or CFL, phase). The CFL
phase is the only phase where both chiral and diquark condensates are simultaneously non-vanishing~\cite{ref:Alford}. In Ref.~\onlinecite{ref:color} it is argued that the transition between the QGP and the CFL phase must be of first order due to active role of the gauge fields.

Needless to say that the system of strongly interacting gluons coupled to dynamical quarks in (3+1) dimensions is much more complicated than a simplified gauge model of correlated electrons in (2+1) dimensions. However, there are some qualitative similarities between these theories as far as the structure of the phase diagram is concerned comparing, say Figure~\ref{fig:phase:2D:section1}~(a) with
Figure~\ref{fig:QCD}. The compact Abelian two-Higgs model may be realized in one of four different phases: Fermi liquid, spin gap, strange metal and superconducting phases which are characterized by pairs
of vanishing or non-vanishing condensates ($b$- or holon, $\Delta$- or spinon-pair condensates,
see also Figure~\ref{fig:Lee:Nagaosa}). The QCD phases can also be characterized by a pair of quark and diquark condensates $(\langle \bar\psi\psi\rangle,\langle \psi\psi\rangle)$. According to Figure~\ref{fig:beta10}, an enhancement of the strength of the phase transition may take place (at sufficiently strong gauge coupling) where both condensates simultaneously change from vanishing to non-vanishing. The same pattern is realized in QCD between the CFL phase and the QGP phase.
The corresponding segment of the phase transition is the result of joining two ``transitions'', the crossover transition between the 2SC color superconductor and the QGP on one hand and the first order phase transition between the 2SC phase and the CFL phase.

Our experience with the cA2HM suggests that such a merging of two weaker transitions -- provided they separate the phase characterized by simultaneously vanishing condensates from the phase with all condensates non-vanishing -- may lead to an enhancement of the order compared to the strength of the individual transition lines. Therefore we expect that at sufficiently large baryon density
the finite-temperature transition between the CFL  phase and the QGP phases should be much stronger than between 2SC and CFL superconducting phases taking place at lower temperatures.

As we have discussed above, the key ingredient of the transition strengthening in the effective model of the strongly correlated electrons is a gauge-boson mediated interaction between the species of the matter fields. As a result, the transition in different channels/species merge and the merger corresponds to a much stronger phase transition. The strengthening effect cannot be seen in QCD phase diagram studies using
the effective NJL model of quark matter (see, for example, Ref.~\onlinecite{Ruster:2005jc}) because that model possesses global symmetries only and the gauge field mediation is obviously absent.
There are, in fact, indications~\cite{ref:color} that in QCD the (thermal) fluctuations of the gauge
fields strengthen the first order phase transition between the QGP and CFL phases, which
nicely matches with our qualitative expectation.

\section{Conclusions}

In this paper we have studied in detail the phase diagram of the $(2+1)D$ Abelian Higgs model with two Higgs fields and one compact gauge field. At weak gauge coupling (large $\beta$) the phase diagram contains two transition lines which run through the whole plane in the hopping parameter
($\kappa_h$, $\kappa_s$)-plane. The transitions are associated with the onset of vortex percolation,
and with the appearances of holon and spinon-pair condensates. The pattern of non-vanishing condensates in different regions of the phase diagram allows to identify the Fermi liquid, spin gap, superconductor and strange metallic phases.

With decreasing $\beta$ above the critical coupling of the gauge-Ising model the strange metallic and Fermi liquid phases become analytically connected at small $\kappa_s$, eventually still being separated by a percolation transition.

At extremely strong coupling (with $\beta$ below the gauge-Ising model transition), however, one of the segments of the transitions disappears together with the spin gap phase, while the difference between the strange metal and the Fermi liquid phases becomes almost invisible since these phases are separated by a smooth crossover in this limit.

The most intriguing effect -- which was first suggested in Ref.~\onlinecite{ref:PRB2006} -- is a strong enhancement of the strength of the phase transition. This enhancement appears in the regime of moderately strong gauge coupling in which the two second order transitions -- corresponding to spinon-pair and
holon condensation lines -- join. The enhancement happens along the common segment of the phase
transition and is reflected in lowering the transition order: two second order transitions become
a single first order transition between the two phases where the two condensates are both vanishing
and both non-vanishing. That strengthening is a result of the activity of the internal gauge field
which couples the dynamics of the spinon-pair and holon condensates.

We suggest that a similar enhancement effect may also be realized in Quantum Chromodynamics at
non-zero temperature and at finite baryon density. At sufficiently large baryon density the finite-temperature transition between the (3-flavor paired) color superconducting phase and the
quark-gluon plasma phases should be much stronger compared with the transition between 2-flavor paired
and 3-flavor paired superconducting phases. The suggested enhancement is emphasized in
Figure~\ref{fig:QCD}.

\begin{acknowledgments}
E.-M.~I. is supported by the DFG Forschergruppe 465
``Gitter-Hadronen-Ph\"anomenologie''.
M.N.~Ch. is supported by the grants RFBR 05-02-16306a, RFBR-DFG 06-02-04010
and by a STINT Institutional grant IG2004-2 025.
M.N.~Ch. wishes to thank I.~Ichinose, T.~Matsui and J.J.M. Verbaarschot
for illuminating discussions. He is grateful to the members of the Department
of Theoretical Physics at Uppsala University for the kind hospitality and
stimulating environment.
\end{acknowledgments}

\end{document}